\ifdefined\blindreview
\documentclass[sigconf, screen, review]{acmart}
\else
\documentclass[conference,hidelinks]{IEEEtran}
\fi

\AtBeginDocument{%
  }

\usepackage{glossaries}
\usepackage{hyperref}
\usepackage[nameinlink]{cleveref}
\usepackage{siunitx}
\usepackage[position=b]{subcaption}
\usepackage{orcidlink}
\usepackage{listings}
\usepackage{multirow}
\usepackage{booktabs}
\usepackage{makecell}
\usepackage{threeparttable}

\glsdisablehyper

\DeclareSIUnit{\ge}{GE}
\DeclareSIUnit{\FLOPS}{FLOPS}

\makeatletter
\@namedef{schnizoname@gp-l}{Schnizo-GP-L}
\@namedef{schnizoname@gp}{Schnizo-GP}
\@namedef{schnizoname@la}{Schnizo-LA}
\@namedef{schnizoname@fp}{Schnizo-FP}
\newcommand{\schnizo}[1]{%
  \ifcsname schnizoname@#1\endcsname
    \csname schnizoname@#1\endcsname
  \else
    \errmessage{Unknown Schnizo variant: #1}%
  \fi
}
\makeatother

\makeatletter
\@namedef{scalarname@gp-l}{Scalar-GP}
\@namedef{scalarname@la}{Scalar-LA}
\newcommand{\scalar}[1]{%
  \ifcsname scalarname@#1\endcsname
    \csname scalarname@#1\endcsname
  \else
    \errmessage{Unknown scalar variant: #1}%
  \fi
}
\makeatother

\newacronym[plural=WANs, firstplural={Wide Area Networks (WANs)}]{wan}{WAN}{Wide Area Network}
\newacronym[plural=WSNs, firstplural={Wireless Sensor Networks (WSNs)}]{wsn}{WSN}{Wireless Sensor Network}
\newacronym{simd}{SIMD}{Single Instruction Multiple Data}
\newacronym{os}{OS}{Operating System}
\newacronym{ble}{BLE}{Bluetooth Low-Energy}
\newacronym{wifi}{Wi-FI}{Wireless Fidelity}
\newacronym[plural=DVS, firstplural={Dynamic Vision Sensors (DVS)}]{dvs}{DVS}{Dynamic Vision Sensor}
\newacronym{ptz}{PTZ}{Pan-Tilt Unit}

\newacronym[plural=FLLs,firstplural=Frequency Locked Loops (FLLs)]{fll}{FLL}{Frequency Locked Loop}
\newacronym{dram}{DRAM}{Dynamic Random Access Memory}
\newacronym{fpu}{FPU}{Floating Point Unit}
\newacronym{rtl}{RTL}{register transfer level}
\newacronym{ff}{FF}{flip-flop}
\newacronym{fpss}{FPSS}{Floating Point Subsystem}
\newacronym{frep}{FREP}{floating point repetition}
\newacronym{dma}{DMA}{Direct Memory Access}
\newacronym{ssr}{SSR}{stream semantic register}
\newacronym{issr}{ISSR}{Indirection Stream Semantic Register}
\newacronym[plural=LUTs, firstplural={Lookup Tables (LUTs)}]{lut}{LUT}{Lookup Table}
\newacronym[plural=FPGAs, firstplural={Field Programmable Gate Arrays (FPGAs)}]{fpga}{FPGA}{Field Programmable Gate Array}
\newacronym{dsp}{DSP}{Digital Signal Processing}
\newacronym{mcu}{MCU}{Microcontroller Unit}
\newacronym{spi}{SPI}{Serial Peripheral Interface}
\newacronym{cpi}{CPI}{Camera Parallel Interface}
\newacronym{rf}{RF}{register file}
\newacronym{prf}{PRF}{physical register file}
\newacronym{fifo}{FIFO}{First-In First-Out Queue}
\newacronym{uart}{UART}{Universal Asynchronous Receiver-Transmitter}
\newacronym{raw}{RAW}{Read After Write}
\newacronym{war}{WAR}{Write After Read}
\newacronym{waw}{WAW}{Write After Write}
\newacronym[plural=ISAs, firstplural={Instruction Set Architectures (ISAs)}]{isa}{ISA}{Instruction Set Architecture}
\newacronym{xbar}{XBAR}{crossbar}
\newacronym[firstplural=Scratch-Pad Memories (SPMs)]{spm}{SPM}{Scratch-Pad Memory}
\newacronym{ipi}{IPI}{Inter-Processor Interrupt}
\newacronym[firstplural=Software-Generated Interrupts (SGIs)]{sgi}{SGI}{Software-Generated Interrupt}
\newacronym{pe}{PE}{processing element}
\newacronym{tcdm}{TCDM}{Tightly-Coupled Data Memory}
\newacronym{lsu}{LSU}{Load-Store Unit}
\newacronym{icache}{I\$}{Instruction Cache}
\newacronym{dcache}{D\$}{Data Cache}
\newacronym{wfi}{WFI}{Wait For Interrupt}
\newacronym{gpc}{GPC}{GPU Processing Cluster}
\newacronym{cpu}{CPU}{Central Processing Unit}
\newacronym{gpu}{GPU}{graphics processing unit}
\newacronym{llc}{LLC}{Last-Level Cache}
\newacronym{sm}{SM}{Streaming Multiprocessor}
\newacronym[firstplural=Networks on Chip (NoCs)]{noc}{NoC}{Network on Chip}
\newacronym{slp}{SLP}{Superscalar Loop Processor}
\newacronym{fu}{FU}{functional unit}
\newacronym{ooo}{OoO}{out-of-order}
\newacronym{ilp}{ILP}{instruction-level parallelism}
\newacronym{dlp}{DLP}{data-level parallelism}
\newacronym{tlp}{TLP}{thread-level parallelism}
\newacronym{ppa}{PPA}{performance, power, area}
\newacronym{zol}{ZOL}{zero-overhead loop}
\newacronym{zolb}{ZOLB}{zero-overhead loop buffer}
\newacronym{dcp}{DCP}{Dataflow Construction Phase}
\newacronym{dcp1}{DCP1}{Dataflow Construction Phase}
\newacronym{dcp2}{DCP2}{Dataflow Construction Phase}
\newacronym{dep}{DEP}{Dataflow Execution Phase}
\newacronym{rs}{RS}{reservation station}
\newacronym[plural={RS entries}, shortplural={RS entries}, longplural={RS entries}, firstplural={RS entries}]{rse}{RS entry}{RS entry}
\newacronym{rmt}{RMT}{register map table}
\newacronym{rob}{ROB}{reorder buffer}
\newacronym{odn}{ODN}{operand delivery network}
\newacronym{cdb}{CDB}{common data bus}
\newacronym{csr}{CSR}{control and status register}
\newacronym{cm}{CM}{constant memory}
\newacronym[plural={CM entries}, shortplural={CM entries}, longplural={CM entries}, firstplural={CM entries}]{cme}{CM entry}{CM entry}

\newacronym{dfg}{DFG}{Data Flow Graph}
\newacronym{lcg}{LCG}{Linear Congruential Generator}
\newacronym{prn}{PRN}{Pseudo-Random Number}

\newacronym{ste}{STE}{Straight-Through-Estimator}

\newacronym[plural=PTUs, firstplural={Pan-Tilt Units}]{ptu}{PTU}{Pan-Tilt Unit}
\newacronym{mdf}{MDF}{Medium-density fibreboard}
\newacronym{cvat}{CVAT}{Computer Vision Annotation Tool}
\newacronym{coco}{COCO}{Common Objects in Context}
\newacronym{soa}{SoA}{state-of-the-art}
\newacronym{sf}{SF}{Sensor Fusion}

\newacronym{dl}{DL}{Deep Learning}
\newacronym{bn}{BN}{Batch Normalization}
\newacronym{FGSM}{FBK}{Fast Gradient Sign Method}
\newacronym{lr}{LR}{Learning Rate}
\newacronym{sgd}{SGD}{Stochastic Gradient Descent}
\newacronym{gd}{GD}{Gradient Descent}
\newacronym{llm}{LLM}{Large Language Model}

\newacronym{sta}{STA}{Static Timing Analysis}

\newacronym[plural=GPIOs, firstplural={General Purpose Inupt Outputs (GPIOs)}]{gpio}{GPIO}{General Purpose Input Output}
\newacronym[plural=LDOs, firstplural={Low Dropout Regulators (LDOs)}]{ldo}{LDO}{Low Dropout Regulator}

\newacronym{inq}{INQ}{Incremental Network Quantization}

\newacronym{CV}{CV}{Computer Vision}
\newacronym{EoT}{EoT}{Expectation over Transformation}
\newacronym{RPN}{RPN}{Region Proposal Network}
\newacronym{TV}{TV}{Total Variation}
\newacronym{NPS}{NPS}{Non-Printability Score}
\newacronym{STN}{STN}{Spatial Transformer Network}
\newacronym{MTCNN}{MTCNN}{Multi-Task Convolutional Neural Network}
\newacronym{YOLO}{YOLO}{You Only Look Once}
\newacronym{SSD}{SSD}{Single Shot Detector}
\newacronym{sota}{SOTA}{state-of-the-art}
\newacronym{NMS}{NMS}{Non-Maximum Suppression}
\newacronym{ic}{IC}{Integrated Circuit}
\newacronym{tcxo}{TCXO}{Temperature Controlled Crystal Oscillator}
\newacronym{jtag}{JTAG}{Joint Test Action Group industry standard}
\newacronym{swd}{SWD}{Serial Wire Debug}
\newacronym{sdio}{SDIO}{Serial Data Input Output}

\newacronym[plural=PCBs, firstplural={Printed Circuit Boards (PCB)}]{pcb}{PCB}{Printed Circuit Board}
\newacronym[plural=ASICs, firstplural={Application Specific Integrated Circuits}]{asic}{ASIC}{Application Specific Integrated Circuit}

\newacronym[plural=BNNs, firstplural={Binary Neural Networks (BNNs)}]{bnn}{BNN}{Binary Neural Network}
\newacronym[plural=NNs, firstplural={Neural Networks}]{nn}{NN}{Neural Network (NNs)}
\newacronym[plural=SCMs, firstplural={Standard Cell Memories (SCMs)}]{scm}{SCM}{Standard Cell Memory}
\newacronym{ann}{ANN}{Artificial Neural Networks}
\newacronym{ml}{ML}{machine learning}
\newacronym{hpc}{HPC}{High-Performance Computing}
\newacronym{ai}{AI}{artificial intelligence}
\newacronym{iot}{IoT}{Internet of Things}
\newacronym{fft}{FFT}{Fast Fourier Transform}
\newacronym[plural=OCUs, firstplural={Output Channel Compute Units (OCUs)}]{ocu}{OCU}{Output Channel Compute Unit}
\newacronym{alu}{ALU}{Arithmetic Logic Unit}
\newacronym{mac}{MAC}{Multiply-Accumulate}
\newacronym[firstplural={systems-on-chip (SoCs)}]{soc}{SoC}{system-on-chip}
\newacronym[firstplural={multi-processor systems-on-chip (MPSoCs)}]{mpsoc}{MPSoC}{multi-processor system-on-chip}

\newacronym{PGD}{PGD}{Projected Gradient Descend}
\newacronym{CW}{CW}{Carlini-Wagner}
\newacronym{OD}{OD}{Object Detection}

\newacronym{rrf}{RRF}{RADAR Repetition Frequency}
\newacronym{nlp}{NLP}{Natural Language Processing}
\newacronym{qam}{QAM}{Quadrature Amplitude Modulation}
\newacronym{rri}{RRI}{RADAR Repetition Interval}
\newacronym{radar}{RADAR}{Radio Detection and Ranging}
\newacronym{loocv}{LOOCV}{Leave-one-out cross validation}

\newacronym{bsp}{BSP}{Board Support Package}
\newacronym{ttn}{TTN}{The Things Network}
\newacronym{wip}{WIP}{Work in Progress}
\newacronym{json}{JSON}{JavaScript Object Notation}
\newacronym{qat}{QAT}{Quantization-Aware Training}

\newacronym{cls}{CLS}{Classification Error}
\newacronym{loc}{LOC}{Localization Error}
\newacronym{bkgd}{BKGD}{Background Error}
\newacronym{roc}{ROC}{Receiver Operating Characteristic}
\newacronym{frr}{FRR}{False Rejection Rate}
\newacronym{eer}{EER}{Equal Error Rate}
\newacronym{snr}{SNR}{Signal-to-Noise Ratio}
\newacronym{flop}{FLOP}{Floating-Point Operation}
\newacronym{fp}{FP}{floating-point}
\newacronym{fps}{FPS}{Frames Per Second}
\newacronym{oi}{OI}{Operational Intensity}
\newacronym{ipc}{IPC}{instructions per cycle}

\newacronym{gsc}{GSC}{Google Speech Commands}
\newacronym{mswc}{MSWC}{Multilingual Spoken Words Corpus}
\newacronym{demand}{DEMAND}{Diverse Environments Multichannel Acoustic Noise Database}

\newacronym[plural=SNNs, firstplural={Spiking Neural Networks (SNNs)}]{snn}{SNN}{Spiking Neural Network}
\newacronym[plural=DNNs, firstplural={Deep Neural Networks (DNNs)}]{dnn}{DNN}{Deep Neural Network}
\newacronym[plural=TCNs,firstplural=Temporal Convolutional Networks]{tcn}{TCN}{Temporal Convolutional Network}
\newacronym[plural=CNNs,firstplural=Convolutional Neural Networks (CNNs)]{cnn}{CNN}{Convolutional Neural Network}
\newacronym[plural=TNNs,firstplural=Ternarized Neural Networks]{tnn}{TNN}{Ternarized Neural Network}
\newacronym{ds-cnn}{DS-CNN}{Depthwise Separable Convolutional Neural Network}
\newacronym{rnn}{RNN}{Recurrent Neural Network}
\newacronym{gcn}{GCN}{Graph Convolutional Network}
\newacronym{mhsa}{MHSA}{Multi-Head Self Attention}
\newacronym{crnn}{CRNN}{Convolutional Recurrent Neural Network}
\newacronym{clca}{CLCA}{Convolutional Linear Cross-Attention}

\newacronym{bf}{BF}{Beamforming}
\newacronym{anc}{ANC}{Active Noise Cancellation}
\newacronym{agc}{AGC}{Automatic Gain Control}
\newacronym{se}{SE}{Speech Enhancement}
\newacronym{mct}{MCT}{Multi-Condition Training}
\newacronym{mcta}{MCTA}{Multi-Condition Training \& Adaptation}
\newacronym{pcen}{PCEN}{Per-Channel Energy Normalization}
\newacronym{mfcc}{MFCC}{Mel-Frequency Cepstral Coefficient}
\newacronym{asr}{ASR}{Automated Speech Recognition}
\newacronym{kws}{KWS}{Keyword Spotting}
\newacronym{odl}{ODL}{On-Device Learning}

\newacronym{nl-kws}{NL-KWS}{Noiseless Keyword Spotting}
\newacronym{na-kws}{NA-KWS}{Noise-Aware Keyword Spotting}
\newacronym{odda}{ODDA}{On-Device Domain Adaptation}
\newacronym{hpm}{HPM}{High-Performance Mode}
\newacronym{lpm}{LPM}{Low-Power Mode}

\makeatletter
\newcommand{\insertnewlines}[1]{%
  \noindent\mbox{}%
  \@tempcnta=#1\relax
  \loop\ifnum\@tempcnta>0
    \\
    \advance\@tempcnta\m@ne
  \repeat
}
\makeatother

\newcommand{\glsinv}[1]{\glsdisp{#1}{\glsentryshort{#1} (\glsentrylong{#1})}}

\newcommand{\ResultSnitchAreaKge}{23}

\newcommand{\ResultScalarAreaKge}{20}

\newcommand{\ResultScalarMulFpuAreaKge}{103}

\newcommand{\ResultSuperscalarSAreaKge}{134}

\newcommand{\ResultSuperscalarLaAreaKge}{264}

\newcommand{\ResultSuperscalarMcAreaKge}{601}

\newcommand{\ResultSuperscalarTrAreaKge}{696}

\newcommand{\ResultSuperscalarGpMAreaKge}{769}

\newcommand{\ResultSuperscalarGpLAreaKge}{1061}
\newcommand{\ResultSnitchAreaRatio}{0.2}
\newcommand{\ResultScalarAreaRatio}{0.2}
\newcommand{\ResultScalarMulFpuAreaRatio}{1.0}
\newcommand{\ResultSuperscalarSAreaRatio}{1.3}
\newcommand{\ResultSuperscalarLaAreaRatio}{2.6}
\newcommand{\ResultSuperscalarMcAreaRatio}{5.8}
\newcommand{\ResultSuperscalarTrAreaRatio}{6.8}
\newcommand{\ResultSuperscalarGpMAreaRatio}{7.5}
\newcommand{\ResultSuperscalarGpLAreaRatio}{10.3}
\newcommand{\ResultSuperscalarSOverheadPct}{30}
\newcommand{\ResultSuperscalarLaOverheadPct}{156}

\newcommand{\ResultSzAxpyAluInsns}{3}
\newcommand{\ResultSzAxpyFpuInsns}{1}
\newcommand{\ResultSzAxpyLsuInsns}{3}
\newcommand{\ResultSzAxpyTotalInsns}{7}
\newcommand{\ResultSzAxpyIdealGpIpc}{7}
\newcommand{\ResultSzAxpyIdealIpc}{7}
\newcommand{\ResultSzDotAluInsns}{2}
\newcommand{\ResultSzDotFpuInsns}{4}
\newcommand{\ResultSzDotLsuInsns}{8}
\newcommand{\ResultSzDotTotalInsns}{14}
\newcommand{\ResultSzDotIdealGpIpc}{3.5}
\newcommand{\ResultSzDotIdealIpc}{3.5}
\newcommand{\ResultExpAluInsns}{22}
\newcommand{\ResultExpFpuInsns}{40}
\newcommand{\ResultExpLsuInsns}{28}
\newcommand{\ResultExpTotalInsns}{90}
\newcommand{\ResultExpIdealGpIpc}{2.2}
\newcommand{\ResultExpIdealIpc}{3.2}
\newcommand{\ResultLogAluInsns}{34}
\newcommand{\ResultLogFpuInsns}{40}
\newcommand{\ResultLogLsuInsns}{16}
\newcommand{\ResultLogTotalInsns}{90}
\newcommand{\ResultLogIdealGpIpc}{2.2}
\newcommand{\ResultLogIdealIpc}{4.5}
\newcommand{\ResultPiLcgAluInsns}{20}
\newcommand{\ResultPiLcgFpuInsns}{28}
\newcommand{\ResultPiLcgLsuInsns}{0}
\newcommand{\ResultPiLcgTotalInsns}{48}
\newcommand{\ResultPiLcgIdealGpIpc}{1.7}
\newcommand{\ResultPiLcgIdealIpc}{3.4}
\newcommand{\ResultPiXoshiropAluInsns}{84}
\newcommand{\ResultPiXoshiropFpuInsns}{28}
\newcommand{\ResultPiXoshiropLsuInsns}{0}
\newcommand{\ResultPiXoshiropTotalInsns}{112}
\newcommand{\ResultPiXoshiropIdealGpIpc}{4}
\newcommand{\ResultPiXoshiropIdealIpc}{4}
\newcommand{\ResultPolyLcgAluInsns}{20}
\newcommand{\ResultPolyLcgFpuInsns}{40}
\newcommand{\ResultPolyLcgLsuInsns}{0}
\newcommand{\ResultPolyLcgTotalInsns}{60}
\newcommand{\ResultPolyLcgIdealGpIpc}{1.5}
\newcommand{\ResultPolyLcgIdealIpc}{3}
\newcommand{\ResultPolyXoshiropAluInsns}{84}
\newcommand{\ResultPolyXoshiropFpuInsns}{40}
\newcommand{\ResultPolyXoshiropLsuInsns}{0}
\newcommand{\ResultPolyXoshiropTotalInsns}{124}
\newcommand{\ResultPolyXoshiropIdealGpIpc}{3.1}
\newcommand{\ResultPolyXoshiropIdealIpc}{4.4}

\newcommand{\ResultGeomeanIdealIpc}{4}
\newcommand{\ResultGeomeanIdealIpcIncrOverGpl}{16}
\newcommand{\ResultNumNonGemmLayers}{547}

\newcommand{\ResultRsAreaOnePortTotSlopeKgePerRse}{1.45}

\newcommand{\ResultRsAreaTwoPortTotSlopeKgePerRse}{1.90}

\newcommand{\ResultRsAreaThreePortTotSlopeKgePerRse}{2.02}

\newcommand{\ResultRsAreaMinRsq}{0.98}

\newcommand{\ResultRsConstantsOnePortTotSlopeKgePerConstant}{0.78}
\newcommand{\ResultRsConstantsTwoPortTotSlopeKgePerConstant}{0.79}
\newcommand{\ResultRsConstantsThreePortTotSlopeKgePerConstant}{0.78}
\newcommand{\ResultRsConstantsMinRsq}{0.99}
\newcommand{\ResultReqXbarOnePortSlopeKgePerRs}{0.94}

\newcommand{\ResultReqXbarTwoPortSlopeKgePerRs}{1.26}

\newcommand{\ResultReqXbarThreePortSlopeKgePerRs}{1.40}

\newcommand{\ResultReqXbarCostPerPortKgePerRs}{0.23}
\newcommand{\ResultReqXbarCostPerPortRsq}{0.95}

\newcommand{\ResultReqXbarSlotsThreePortSlopeKgePerRse}{0.15}
\newcommand{\ResultReqXbarSlotsThreePortRsq}{0.99}

\newcommand{\ResultRspXbarOnePortSlopeKgePerRs}{2.58}

\newcommand{\ResultRspXbarTwoPortSlopeKgePerRs}{5.33}

\newcommand{\ResultRspXbarThreePortSlopeKgePerRs}{8.30}

\newcommand{\ResultRspXbarMinRsq}{0.98}
\newcommand{\ResultRspXbarCostPerPortKgePerRs}{2.86}
\newcommand{\ResultRspXbarCostPerPortRsq}{0.99}

\newcommand{\ResultGeomeanFractionOfIdealIpcExclDiv}{92}
\newcommand{\ResultGeomeanFpuUtilExclDiv}{92}
\newcommand{\ResultAsymptoticGeomeanFractionOfIdealIpcExclDiv}{95}
\newcommand{\ResultAsymptoticGeomeanFpuUtilExclDiv}{95}
\newcommand{\ResultGeomeanFractionOfFcIpcWithOnePort}{79}
\newcommand{\ResultGeomeanFractionOfFcIpcWithTwoPorts}{97}
\newcommand{\ResultGeomeanFractionOfFcIpcWithThreePorts}{100}
\newcommand{\ResultLaGeomeanSpeedup}{4.4}
\newcommand{\ResultLaGeomeanAreaEfficiency}{16.5}
\newcommand{\ResultLaGeomeanAreaEfficiencyImpr}{72}
\newcommand{\ResultLaGeomeanAreaEfficiencyImprCluster}{195}

\newcommand{\ResultMaxGeomeanSpeedupExclLa}{2.6}
\newcommand{\ResultSpatzLaGeomeanAreaEfficiencyDecrease}{46}
\newcommand{\ResultSpatzVrfAreaToSchnizoLaAreaRatio}{80}

\begin{document}

\title{Superscalar Loop Processors}

\ifdefined\blindreview
\else


\author{
    \IEEEauthorblockN{Luca Colagrande}
    \IEEEauthorblockA{
        \textit{Integrated Systems Laboratory (IIS)} \\
        \textit{ETH Zurich}\\
        Zurich, Switzerland \\
        colluca@iis.ee.ethz.ch\,\orcidlink{0000-0002-7986-1975}
    }
    \and
    \IEEEauthorblockN{Pascal Etterli}
    \IEEEauthorblockA{
        \textit{D-ITET} \\
        \textit{ETH Zurich}\\
        Zurich, Switzerland \\
        petterli@student.ethz.ch\,\orcidlink{0009-0002-5630-5738}
    }
    \and
    \IEEEauthorblockN{Luca Benini}
    \IEEEauthorblockA{
        \textit{Integrated Systems Laboratory (IIS)} \\
        \textit{ETH Zurich}\\
        Zurich, Switzerland \\
        lbenini@iis.ee.ethz.ch\,\orcidlink{0000-0001-8068-3806}
    }
}
\fi


\maketitle

\begin{abstract}

To sustain the computational demands of modern ML and HPC workloads, programmable accelerators integrate large arrays of processing elements (PEs), typically built around single-issue in-order cores to maximize integration density and energy efficiency.
As a result, each PE's instruction throughput is upper-bounded at one instruction per cycle (IPC), often suffering from significant underutilization of its functional units (FUs).
Conventional superscalar processors can improve FU utilization by increasing IPC, but they typically incur substantial area and energy overheads.
This work proposes the Superscalar Loop Processor (SLP) architecture, a lightweight approach to superscalar execution that targets regular, compute-intensive loop kernels.
We present the SLP execution model, which exploits the static structure of these workloads to enable efficient multiple-issue execution without the complexity of dynamic out-of-order scheduling.
We demonstrate the SLP architecture through an open-source prototype implementation, Schnizo, developed as an enhancement of a tiny RV32I core with floating-point support.
Across several ML and HPC microkernels, Schnizo achieves substantially higher IPC and FPU utilization than the scalar baseline while retaining the efficiency of lightweight in-order designs.
These results demonstrate that SLPs can provide an efficient architectural solution for programmable accelerators targeting regular, compute-intensive workloads.
\end{abstract}


\glsunset{isa}
\glsunset{fifo}
\glsunset{raw}
\glsunset{war}
\glsunset{waw}
\glsunset{simd}
\glsunset{fpu}
\glsunset{dcp1}
\glsunset{dcp2}
\glsunset{csr}
\glsunset{hpc}
\glsunset{cm}
\glsunset{rtl}
\glsunset{dlp}
\glsunset{rse}
\glsunset{cme}

\lstset{
    language=C,                        
    basicstyle=\ttfamily\footnotesize, 
    literate={~}{{\fontfamily{ptm}\selectfont \textasciitilde}}1
}

\section{Introduction}
\label{sec:introduction}

The growth of \gls{ml} workloads has created an unprecedented demand for computational capacity \cite{openai_compute}.
Empirical studies show that model quality improves predictably with additional compute \cite{kaplan2020scaling,hoffmann2022compute}, driving rapid expansion of data centers and specialized accelerators.
Recent analyses report that frontier-model training compute continues to grow at roughly 4-5$\times$ per year \cite{sevilla2024training} and inference is scaling even faster due to the explosive increase in token generation linked to agentic workloads \cite{openai_enterprise_ai_2025}.

To meet this demand, modern accelerators integrate an ever-increasing number of arithmetic or \glspl{fu} \cite{jung2024hammerblade, abts2020thinkfast, firoozshahian2023mtia, choquette2023blackwell, coburn2025mtia2, prabhakar2024sn40l, vasiljevic2024blackhole, lie2024wse3, rico2024xdna, scheffler2025occamy, liao2021ascend}.
Area efficiency (\si[per-mode=symbol]{\FLOPS\per\milli\meter\squared}) has thus become a primary design objective, as it enables designers to pack more compute resources onto a chip, under a fixed silicon budget.
At the same time, energy has become a constraint on how much compute infrastructure can be deployed, with hyperscalers investing in dedicated power plants to scale their data centers \cite{hai2025_ai_index}.
As such, energy efficiency (\si[per-mode=symbol]{\FLOPS\per\watt}) has also become a critical design goal, as it enables deploying more compute infrastructure within a fixed power budget.

\begin{figure}[t]
  \centering
  \begin{subfigure}{0.7\columnwidth}
    \centering
    \includegraphics[width=\textwidth]{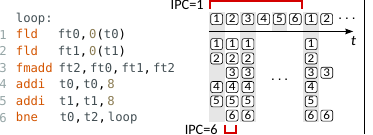}
    \caption{}
    \label{fig:dotp_naive}
  \end{subfigure}
  \hfill
  \begin{subfigure}{0.28\columnwidth}
      \centering
      \includegraphics[width=\textwidth]{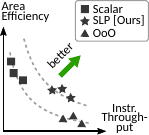}
      \caption{}
      \label{fig:design_space}
  \end{subfigure}
  \caption{(a) Dot product loop and sample schedules on idealized scalar (top, 17\% FPU utilization) and superscalar (bottom, 100\% FPU utilization) cores. (b) Architectural design space. Contour lines indicate equal $\text{Area Eff.} \times \text{IPC}$.}
\end{figure}

While integrating more \glspl{fu} can effectively boost peak performance, attaining this performance requires sustaining a high \gls{fu} utilization.
As most kernels involve a mix of compute and non-compute instructions \cite{talla2003}, e.g. for data movement and control operations, compute utilization critically depends on instruction throughput.
This is evident in the dot product kernel illustrated in \Cref{fig:dotp_naive}.
As only one in six instructions performs a \gls{fp} arithmetic operation, a processor with an \glsinv{ipc} of 1 can achieve only 17\% of its peak \gls{fp} performance.
The following equations capture these relationships:
\begin{align*}
  &\quad\text{Peak Perf.} \,[\mathit{FLOPS}] = \min(\\
  &\quad\qquad\text{Area}  \,[\mathit{mm^2}] \times \text{Area Eff.}   \,[\mathit{FLOPS/mm^2}],\\
  &\quad\qquad\text{Power} \,[\mathit{W}]    \times \text{Energy Eff.} \,[\mathit{FLOPS/W}])\\
  &\quad\text{Utilization} = \text{Op. Density} \,\left[\frac{\mathit{op}}{\mathit{ins}}\right] \times
                       \text{IPC} \,\left[\frac{\mathit{ins}}{\mathit{cyc}}\right] \div
                       1 \left[\frac{\mathit{cyc}}{\mathit{op}}\right]\\
  &\quad\text{Attained Perf.} \,[\mathit{FLOPS}] = \text{Peak Perf.} \,[\mathit{FLOPS}] \times \text{Util.}
\end{align*}

While scalar in-order cores offer high area and energy efficiency \cite{zaruba2021snitch, scheffler2025occamy, baas2007asap, bohnenstiehl2016, jung2024hammerblade}, they achieve limited instruction throughput (\gls{ipc} $\leq$ 1).
On the other hand, superscalar \gls{ooo} cores provide excellent instruction throughput across a wide range of applications, but require large structures such as reorder buffers and dynamic scheduling logic, which incur substantial area and energy costs \cite{palacharla1997complexity, kucuk2004, folegnani2001, huang2002}.
\Cref{fig:design_space} illustrates this design space.

In practice, many modern programmable accelerators, including \glspl{gpu}, rely on large arrays of relatively simple execution units and in-order scheduling structures to exploit data- (DLP) or \gls{tlp}, rather than on complex \gls{ooo} cores or scheduling structures to extract \gls{ilp} within individual instruction streams \cite{scheffler2025occamy, jung2024hammerblade, choquette2023blackwell, miyajima2025}.
This suggests that the area- and energy-efficiency of simple replication-friendly designs outweighs the \gls{ipc} advantage of traditional superscalar pipelines on the regular, computationally-intensive kernels found in \gls{ml} workloads \cite{feng2023tensorir, karami2025nongemm, vasilache2018}.

\begin{figure}[t!]
    \centering
    \begin{subfigure}{0.32\columnwidth}
        \centering
        \includegraphics[width=\textwidth]{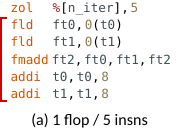}
        \phantomcaption
        \label{fig:zol}
    \end{subfigure}
    \hfill
    \begin{subfigure}{0.32\columnwidth}
        \centering
        \includegraphics[width=\textwidth]{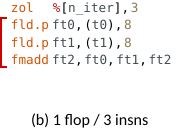}
        \phantomcaption
        \label{fig:post_increment}
    \end{subfigure}
    \hfill
    \begin{subfigure}{0.32\columnwidth}
        \centering
        \includegraphics[width=\textwidth]{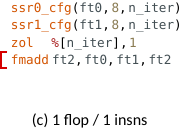}
        \phantomcaption
        \label{fig:ssr}
    \end{subfigure}
    \caption{
      Dot product kernel leveraging several extensions: (a) ZOL, (b) ZOL and post-increment loads, (c) ZOL and SSRs.
      Highlighted in red is the loop body, and annotated under the loop is the FPU operation density within the loop.
    }
\end{figure}

A complementary way to improve \gls{fu} utilization is to increase the operation density, i.e., reduce the number of instructions required per useful operation.
This approach uses custom instructions to fuse common instruction sequences \cite{gautschi2017pulp, cui2023, galuzzi2011}.
Notable examples include \glspl{zol} (or hardware loops) \cite{bajwa1997, uh1999, tsao2003, talla2003, kavvadias2008, kavvadias2010, wei2010, vadivel2017, gautschi2017pulp, colagrande2025gemm}, post-increment load/store instructions \cite{fan2025, lee1987, gautschi2017pulp} and \glspl{ssr} \cite{schuiki2021ssr, scheffler2023sssr}, which transform the dot product code as shown in \Cref{fig:zol,fig:post_increment,fig:ssr}.
Although these techniques are often adopted in in-order cores to compensate for limited instruction throughput, they remain orthogonal to the tradeoff between efficiency and instruction throughput.

This tradeoff leads to a central question: \emph{can we design a lightweight superscalar processor that provides high instruction throughput while retaining the efficiency benefits of simple in-order designs?}
More specifically, can we trade the flexibility of traditional \gls{ooo} superscalar processors for a more efficient design that still achieves high throughput on regular, compute-intensive kernels?

In this work, we propose \emph{\glspl{slp}}, a lightweight processor architecture that specializes superscalar execution for regular loop kernels.
Rather than supporting general-purpose \gls{ooo} execution, \glspl{slp} extract \gls{ilp} exclusively within regular loop bodies.
By constraining the scope of superscalar execution, \glspl{slp} achieve high \gls{fu} utilization without incurring the overhead of \gls{ooo} scheduling, bridging the gap between scalar efficiency and superscalar throughput for regular, compute-intensive workloads.
We envision \glspl{slp} as an enhancement to the \glspl{pe} of high-throughput programmable accelerators, such as \glspl{gpu} and many-cores, where execution is dominated by regular loop kernels.
They are not meant to replace \gls{ooo} cores, which remain essential for latency-sensitive and irregular workloads.

This paper makes the following contributions:
\begin{enumerate}

\item We provide a critical analysis of earlier pseudo-dual-issue execution methods, identifying their potential and limitations and formulating the design challenges to be addressed.

\item Based on these insights, we introduce the SLP architecture, a lightweight multiple-issue execution method for regular loop kernels, and present an implementation-agnostic algorithm for its operation.

\item We implement the \gls{slp} architecture in Schnizo%
\footnote{We release Schnizo as open source: %
\ifdefined\blindreview
https://hidden-for-double-blind-review.com
\else
\url{https://github.com/colluca/schnizo}.
\fi
}%
, an \gls{rtl} prototype developed as an extension of a lightweight RV32I core, detailing the key hardware structures and control mechanisms required to support its execution model.

\item We characterize Schnizo's scalability across key architectural parameters, and assess its performance on a diverse set of microkernels. On simple linear algebra kernels, Schnizo achieves a geometric mean \ResultLaGeomeanSpeedup$\times$ speedup over scalar execution and \ResultLaGeomeanAreaEfficiencyImpr\% higher area efficiency than a scalar design.

\end{enumerate}

Each contribution is developed in a dedicated section, with \Crefrange{sec:background}{sec:results} covering contributions 1 through 4, respectively.


\section{Background}
\label{sec:background}

\begin{figure}[t!]
  \centering
  \includegraphics[width=\columnwidth]{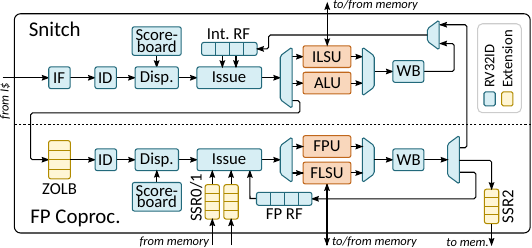}
  \caption{Architecture of Snitch and its FP coprocessor.}
  \label{fig:snitch}
\end{figure}

We base our work on the open-source Snitch processor \cite{zaruba2021snitch}, represented in \Cref{fig:snitch}.
Snitch is a tiny RV32I core with an in-order, single-issue and single-stage pipeline, optionally coupled to a 4-stage double-precision \gls{fp} coprocessor.
\gls{fp} instructions are fetched by the Snitch core and offloaded to the \gls{fp} coprocessor.
A scoreboard is used to track in-flight coprocessor instructions, providing limited latency tolerance at low cost.
Despite its single-stage pipeline, implementation results in GlobalFoundries' \SI{22}{\nano\meter} FDX technology \cite{zaruba2021snitch} show that Snitch achieves a \SI{1}{\giga\hertz} clock frequency in typical operating conditions.
With a footprint of only \SI{21}{\kilo\ge} (kilo gate equivalents), it occupies the extreme end of the design space in \Cref{fig:design_space}, with high efficiency, but limited instruction throughput.

High \gls{fu} utilization is restored through the operation density knob.
Two \gls{isa} extensions are dedicated to this purpose: \glspl{ssr} \cite{schuiki2021ssr} and the \gls{frep} extension \cite{zaruba2021snitch}.
As shown in \Cref{fig:ssr}, \glspl{ssr} eliminate load, store and address generation instructions in the hot loop, by programming all memory accesses once (in compressed form), outside the loop.
This applies to regular memory access patterns (affine streams) only and is achieved by streaming operands through a dedicated prefetcher, while temporarily altering architectural register semantics (for \lstinline{ft0} and \lstinline{ft1} in our example).

The \gls{frep} extension, on the other hand, eliminates branch instructions by implementing a \gls{zolb} \cite{bajwa1997, uh1999, tsao2003, vadivel2017, wei2010, colagrande2025gemm, zaruba2021snitch} which can cache and loop over short sequences of \gls{fp} instructions.
Differently from prior designs, the \gls{zolb} is implemented in the \gls{fp} coprocessor, allowing it to independently issue instructions from the buffer, while the Snitch core is free to fetch and execute other instructions from the instruction cache; a novel scheme its authors refer to as \emph{pseudo-dual issue} execution.

\begin{figure}[t!]
  \centering
  \includegraphics[width=\columnwidth]{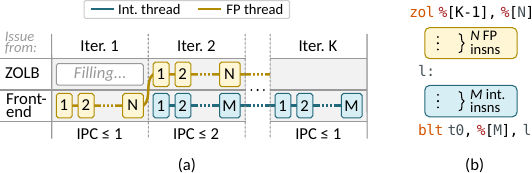}

  \begingroup
    \phantomsubcaption
    \label{fig:frep_schedule}
    \phantomsubcaption
    \label{fig:frep_programming_model}
  \endgroup
  \caption{(a) FREP schedule and (b) programming model.}
\end{figure}

As shown in \Cref{fig:frep_schedule}, this scheme effectively creates two separate, independent ``threads'' of instructions, asynchronously executing across the two decoupled domains, and each maintaining their own architectural state: the \gls{fp} thread state comprises of the \gls{fp} \gls{rf} and the \gls{zolb} state, while the integer thread state comprises of the integer \gls{rf} and the program counter.
However, for a single thread of integer and \gls{fp} instructions to be separable into said independent threads of integer-only and \gls{fp}-only instructions, there can be no dependencies between the integer and \gls{fp} instructions involved.
This excludes all \gls{fp} conversion, compare and move instructions, as well as \gls{fp} loads and stores, thereby severely limiting the applicability of the approach.

Colagrande and Benini \cite{colagrande2025copift} identified this issue and proposed COPIFT, a software methodology that enables pseudo-dual-issue execution even in the presence of dependencies. It does so by enabling communication between the two threads through the L1 memory and by increasing the dependency distance via tiling and software pipelining transformations. 
While effective, the COPIFT methodology introduces a significantly complex programming model as well as non-negligible pipeline fill and drain overheads.
In a follow-up work, they proposed COPIFTv2 \cite{colagrande2026copiftv2}, which refines this scheme by enabling fine-grained communication and synchronization between the two threads through dedicated \glspl{fifo}.
Although this design eliminates much of the complexity of the original approach, it still requires the programmer to explicitly manage communication between the two threads, either through modified \gls{ssr}-like register semantics or through explicit push/pop instructions, which incur additional overhead.

Despite its limitations, the pseudo-dual issue model is appealing because it exposes \gls{ilp} with minimal hardware overhead, preserving the area and energy efficiency of a single-issue in-order design.
It relies on a \emph{single-issue} instruction frontend that dispatches instructions \emph{in order} to the appropriate \glspl{fu}, incrementally setting up superscalar execution.
It then exploits the invariant nature of loops to realize superscalar benefits across successive iterations without additional hardware complexity.
As such, it pioneers an approach to superscalar execution that we set out to extend in this work.

The current pseudo-dual-issue model has two main limitations:
\begin{enumerate}
\item Originally conceived as a specialized mechanism for overlapping integer and \gls{fp} execution, it does not readily generalize to multiple \glspl{fu} or to \glspl{fu} without a dedicated architectural \gls{rf}.
For instance, it cannot directly parallelize integer arithmetic and integer load/store operations.
Moreover, because only the \gls{fp} domain implements \glspl{zol}, integer instructions must still be iterated explicitly in software, leading to the awkward programming model shown in \Cref{fig:frep_programming_model}.
\item Dependencies across threads require explicit communication and synchronization.
As a result, extracting \gls{ilp} requires the programmer or compiler to statically partition a loop body into multiple communicating instruction streams and to expose the required communication and synchronization mechanisms in the \gls{isa}.
The resulting code is thus tied to a specific machine configuration (\gls{fu} counts, latencies, and interconnect characteristics) hindering performance portability or binary compatibility across implementations.
\end{enumerate}

To address these limitations, we set out to:
\begin{enumerate}
\item Generalize the pseudo-dual issue model into a native \emph{multiple-issue} execution model for loops that supports arbitrary \glspl{fu} and exposes a coherent programming model.
\item Handle dependencies between concurrently executing instruction streams dynamically in hardware, transparently to software and with zero instruction overhead.
\end{enumerate}

In the next section, we show how the \gls{slp} architecture addresses these challenges.

\section{Architecture}
\label{sec:architecture}

The \gls{slp} architecture builds on the key idea of leveraging single-issue in-order execution in the first iteration of a loop to incrementally set up the \glspl{fu} for superscalar execution.
The \glspl{fu} then replay subsequent iterations largely asynchronously.
In the following, we refer to the first phase as the \gls{dcp} and the second as the \gls{dep}.

To eliminate the awkward programming model of the pseudo-dual issue approach and generalize it to arbitrary multiple issue, we move away from Snitch's asymmetric processor-coprocessor design.
Instead, all \glspl{fu} are treated uniformly, and each is equipped with its own \gls{zolb}.
The programmer or compiler defines a single unified hardware loop, as in \Cref{fig:zol}.
During the first iteration (\gls{dcp}), the \gls{slp}'s frontend fetches the loop body instructions one by one and dispatches each instruction to the appropriate \gls{fu}, where it is buffered.
Unlike Snitch, the frontend is not tied to any specific \gls{fu}.
In subsequent iterations, the frontend remains idle, and each \gls{fu} independently replays its buffered instructions using dedicated \gls{zol} logic.
Once all \glspl{fu} complete all loop iterations, the hardware loop terminates and the frontend resumes normal scalar execution.

The frontend and \gls{zol} logic are relatively simple when each \gls{fu} is guaranteed to access a disjoint subset of architectural registers, as in Snitch's pseudo-dual issue model.
Under this assumption, the instruction streams dispatched to different \glspl{fu} are independent and can therefore execute fully asynchronously.
However, this assumption becomes too restrictive when scaling to multiple \glspl{fu} and limits its applicability to realistic workloads.
To remove this constraint, the architecture must at least support true (\gls{raw}) dependencies between instructions assigned to different \glspl{fu}.
This requires detecting such dependencies and synchronizing the \glspl{fu} accordingly.
The same issue arises for false (\gls{waw}, \gls{war}) dependencies, unless they are eliminated through register renaming.
In that case, the architecture additionally requires a mechanism to commit renamed, or physical-register, state back to the architectural state.

To this end, we extend the frontend to extract dependency information during \gls{dcp} and record it directly in the \gls{zolb} entry of each instruction.
During \gls{dep}, the \glspl{fu} use this information to dynamically determine when instructions become ready for execution.
In this sense, execution during \gls{dep} follows a dataflow-like model that is established during \gls{dcp}.

For the dataflow construction logic, we draw heavily from conventional \gls{ooo} superscalar processors.
The key simplification relative to a conventional design is that this logic is active only during \gls{dcp} and needs to process at most one instruction per cycle, since the issue width in this phase is one.
The main advantage of \glspl{slp} follows directly from this observation: by exploiting the invariant structure of loops, where the dataflow graph remains unchanged across iterations, \glspl{slp} can decouple the issue width used for dataflow construction from the issue width realized during execution.

To illustrate the general operation of the \gls{slp} architecture, we present an implementation-agnostic \gls{slp} approach based on Tomasulo's algorithm \cite{tomasulo1967}.
In this formulation, the \gls{zolb} associated with each \gls{fu} serves as a \gls{rs} for loop body instructions.
The \gls{slp} frontend is augmented with a \gls{rmt} that tracks the latest producer tag for each architectural register.
When an instruction is dispatched during \gls{dcp}, the frontend reads the producer tags of its source operands from the \gls{rmt} and updates the destination register entry accordingly, following Tomasulo's dependency-tracking scheme.
The source operand tags are then recorded with the instruction in the corresponding \glspl{rs}.
Once the first loop iteration completes, the frontend signals the \glspl{fu} to enter \gls{dep}.
During \gls{dep}, all \glspl{rs} can issue instructions independently and in parallel.
Within an \gls{rs}, instructions need not issue strictly in program order: any instruction may issue as soon as its dependencies, as recorded during \gls{dcp}, are satisfied.
However, each \gls{rs} must track the loop iteration index for all of its instructions.

The iteration index is required for two reasons.
First, it serves the original purpose required by the \gls{zol} logic, namely to determine when all loop iterations have completed.
Second, it is needed to preserve false dependencies.
In Tomasulo's algorithm, each \gls{rse} is assigned a unique tag.
This guarantees that every in-flight instruction within an iteration is uniquely identified, preventing false dependency violations.
In our adaptation, by contrast, the same \gls{rse} issues the same instruction across multiple iterations, and therefore does not uniquely identify a single in-flight instruction.
Reusing a tag in a subsequent loop iteration requires ensuring that the previous value associated with that tag is no longer needed, that is, that all reads of that value have completed.
This condition is trivially satisfied in the absence of loop-carried dependencies, that is, when all values produced in a given iteration are fully consumed within that same iteration.
In this case, it is sufficient to synchronize all \glspl{rs} at the end of each iteration, thereby preventing instructions from different iterations from being issued concurrently.
The frontend can aggregate the iteration indices from all \glspl{rs}, determine when an iteration has completed globally, and then advance all \glspl{rs} to the next iteration.
This approach, however, does not support loop-carried dependencies and can exploit \gls{ilp} only within a single iteration.

Before addressing this limitation, we note that the same principles can be applied to modern superscalar designs based on a physical \gls{rf} \cite{yeager1996, kessler1999, balasubramonian2001, tseng2003}, by assigning physical registers during \gls{dcp} and preserving these assignments across \gls{dep} iterations.
The same restriction as in the Tomasulo-based formulation applies: a physical register cannot be implicitly reassigned to an instruction issue in a subsequent iteration until all reads of its previous value have completed.
This condition can be enforced by disallowing loop-carried dependencies and by preventing instructions from different iterations from being simultaneously in flight.

This policy can be viewed as a coarse-grained physical register reclamation strategy.
In conventional free-list-based designs, a physical register is typically reclaimed when the next instruction writing the same architectural register commits, thereby superseding the previous mapping \cite{yeager1996, sohi1995}.
In contrast, our approach commits at the granularity of an entire loop iteration.
Once all instructions in the current iteration have committed, all physical registers associated with the previous iteration are reclaimed at once, allowing the next iteration to begin.
As noted above, this strategy requires only a single counter per \gls{rs} and an aggregation mechanism in the frontend.
A more fine-grained physical register reclamation strategy, for example based on a free-list or reference counting scheme \cite{moudgill1993, akkary2003, battle2012}, would allow instructions from different iterations to be in flight simultaneously and would therefore enable support for loop-carried dependencies.
We demonstrate such an implementation in \Cref{sec:rs}.

In this section, we presented an implementation-agnostic algorithm to illustrate the high-level operation of the \gls{slp} architecture and to highlight the similarities and differences with conventional superscalar designs.
This discussion is by no means exhaustive, but is intended to convey the key principles of the approach.
In the next section, we present Schnizo and examine the main design challenges of a concrete \gls{slp} implementation in greater detail.
\section{Implementation}
\label{sec:implementation}

\begin{figure}[t!]
  \centering
  \includegraphics[width=\columnwidth]{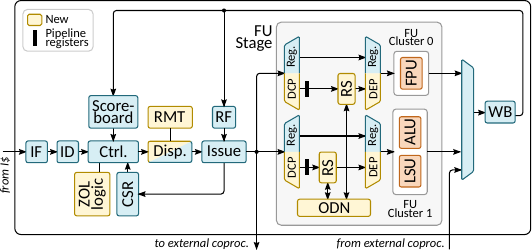}
  \caption{Architecture of a two-\gls{fu}-cluster Schnizo core.}
  \label{fig:schnizo}
\end{figure}

\Cref{fig:schnizo} illustrates the high-level microarchitecture of Schnizo.\footnote{
  The name Schnizo is a portmanteau of schizo- (Ancient Greek for ``to split'') and Snitch: metaphorically, the \gls{slp} extension splits a single ``thread'' of instructions into multiple parallel threads.
}
As in Snitch, Schnizo's scalar pipeline implements a tiny RV32I core with a single-stage, in-order, single-issue design, which can optionally integrate Snitch's 4-stage double-precision \gls{simd}-capable \gls{fpu} as a tightly-coupled \gls{fu}.
Schnizo further inherits Snitch's lightweight scoreboard-based latency-tolerance mechanism.
Finally, we generalize Snitch's \gls{zol} support by moving the loop-control logic into the shared instruction frontend, allowing \glspl{zol} to contain arbitrary instructions rather than only \gls{fp} instructions.

\begin{figure}[t!]
  \centering
  \includegraphics[width=\columnwidth]{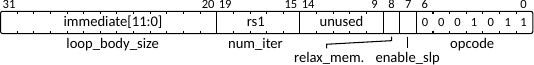}
  \caption{ZOL instruction encoding.}
  \label{fig:zol_instruction}
\end{figure}

On top of this, we implement the \gls{slp} extension, highlighted in yellow in \Cref{fig:schnizo}.
The \gls{slp} extension can be disabled at design time, enabling the same design to be configured as a simple Snitch-compatible scalar core, and providing a clean baseline for comparison.
Thanks to the pipeline registers at the issue interface cutting combinational paths, the extension is also minimally intrusive when enabled, i.e., it does not significantly affect the timing of the scalar pipeline and regular or \gls{zol} execution.
In addition, we support both \gls{slp} and \gls{zol} execution modes through a unified mechanism.
As shown in \Cref{fig:zol_instruction}, a 1-bit \lstinline{enable_slp} flag is added to the \gls{zol} instruction to select whether a loop should be executed in \gls{slp} or in \gls{zol} mode, on an individual basis.
The \gls{zol} mode also serves as a fallback when a loop cannot execute in \gls{slp} mode, for example because of resource constraints or unsupported instruction patterns in the loop body.
Concrete cases are presented in \Cref{sec:limitations}.

The \gls{slp} extension is implemented through four main hardware components, shown in \Cref{fig:schnizo}: \glspl{rs}, an \gls{odn}, a \gls{zol} controller, and \gls{slp}-specific extensions to the dispatch stage for dependency tracking and register renaming.
We describe each component in detail in the following sections.

\subsection{Reservation Station}
\label{sec:rs}

The \gls{rs} is the core of the dataflow execution logic.
Each \gls{rs} is associated with a disjoint set of \glspl{fu}, which we refer to as an \gls{fu} cluster.
The \gls{fu} cluster defines the granularity at which \gls{ilp} is extracted: each \gls{fu} cluster can issue at most one instruction per cycle, but separate \gls{fu} clusters can issue in parallel.
Grouping multiple \glspl{fu} into a single cluster can be beneficial when the cost of a separate cluster per \gls{fu}, with its own \gls{rs} and \gls{odn} ports, outweighs the benefit of parallel execution across the \glspl{fu}.
This is relevant for \glspl{fu} executing different classes of instructions, such as an \glsinv{lsu} and an \glsinv{alu}.
By contrast, there is no benefit in grouping multiple \glspl{fu} of the same type within a cluster, since the same functionality would already be provided by a single \gls{fu}.

\begin{figure}[t!]
  \centering
  \includegraphics[width=\columnwidth]{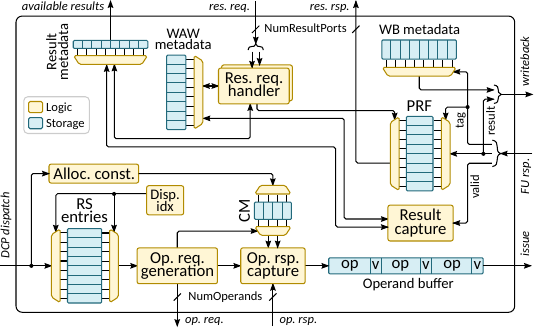}
  \caption{Architecture of the \acrfull{rs}.}
  \label{fig:rs}
\end{figure}

\Cref{fig:rs} illustrates the architecture of the \gls{rs}.
As described in \Cref{sec:architecture}, the \gls{rs} in an \gls{slp} serves two purposes: it caches the loop-body instructions assigned to its associated \glspl{fu} and schedules their execution.
It can therefore store up to one loop body instruction per \gls{rse}.
To reduce implementation complexity, Schnizo restricts scheduling within an \gls{rs} to in-order issue, thereby limiting \gls{ilp} to inter-\gls{rs} parallelism.
As a result, only one instruction within an \gls{rs} is eligible for dispatch at any given time.
This instruction is tracked by the dispatch index (\lstinline{disp_idx}), which advances whenever an instruction is dispatched.

For an instruction to be issued%
\footnote{In the \gls{rs}, or during \gls{dep}, we refer to ``dispatch'' as the process of selecting an instruction for execution, and ``issue'' as the process of forwarding it to the actual \gls{fu} after having fetched all its source operands. In the frontend, or during DCP, we use the terms interchangeably, as the distinction is not relevant to the treatment in this work.}
by the \gls{rs}, its source operands must be available.
The \gls{rs} is therefore responsible for generating the requests needed to retrieve the operands of the selected instruction.
Our design adopts a distributed \gls{prf} organization \cite{kessler1999, farkas1997, palacharla1997complexity, sohi1995multiscalar, smith2001}, in which the \glspl{rs} store the results of the loop-body instructions they hold.
This organization is aligned with the targeted dataflow execution model: results remain close to the producing \glspl{fu} and are forwarded through the \gls{odn} directly to consumers, rather than being written back to and later read from a centralized \gls{prf}.
We further distinguish between static and dynamic operands.
Static operands comprise immediates and architectural register values that are not modified within the loop and therefore remain constant throughout execution.
Dynamic operands comprise values produced by loop-body instructions and are therefore updated during loop execution.

Each \gls{rs} includes a private ``constant memory'' (\gls{cm}) to store static operands.
We support instructions with up to two static operands, served through two read ports into the \gls{cm}.
In addition, every \gls{rs} contains a \gls{prf} with one entry per \gls{rse} to store the results produced by its instructions.
The \gls{prf} has a single write port, which is sufficient to sustain the average result-production rate of the associated \gls{fu} cluster given the maximum issue rate of one instruction per cycle per \gls{rs}.
To support \glspl{fu} that may return results out of order, such as a variable-latency \gls{fpu}, a tag is sent to the \gls{fu} together with the instruction at issue time.
This tag uniquely identifies the \gls{prf} entry associated with the instruction and must be returned by the \gls{fu} alongside the result, allowing the \gls{rs} to determine which entry must capture the result.
The \gls{prf} provides a design-time configurable number of read ports, matching the number of incoming result requests from the \gls{odn} that it can serve per cycle.
Each read port is paired with a dedicated \lstinline{res_req_handler} module to serve these requests.
In \Cref{sec:scalability,sec:tradeoffs}, we will evaluate the impact of the number of read ports on area and performance.

When an instruction is selected for dispatch, requests for static operands are sent to the \gls{cm}, while requests for dynamic operands are placed on the \gls{odn}.
Operand requests on the \gls{odn} are arbitrated and served independently, so the source operands requested by an \gls{rs} may return at different times.
To handle this, the \gls{rs} includes dedicated registers to buffer the returned operands and track their readiness information until all operands are available, at which point the instruction can be issued to the corresponding \gls{fu}.

Because the physical and architectural \glspl{rf} are decoupled, a mechanism is needed to ensure that the architectural register state is consistent at loop completion.
To this end, when an instruction result is captured by the \gls{rs} in its final loop iteration, it writes the result both to the \gls{prf} and back to the architectural \gls{rf}.
To preserve \gls{waw} dependencies, only the last loop-body instruction in program order that writes a given architectural register is permitted to write back to the architectural \gls{rf}.
The \gls{rs} stores a flag (\lstinline{do_writeback}) for each instruction, indicating whether this instruction is the final writer of its architectural destination within the loop and if it should therefore write back to architectural \gls{rf} in the last loop iteration.
This information is recorded during \gls{dcp}, as described in \Cref{sec:controller}.

\begin{figure}[t!]
  \centering
  \includegraphics[width=\columnwidth]{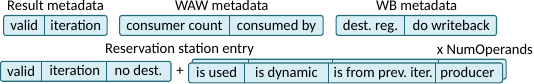}
  \caption{\gls{rs} metadata.}
  \label{fig:metadata}
\end{figure}

As mentioned in \Cref{sec:architecture}, supporting loop-carried dependencies requires a more fine-grained physical register reclamation strategy than the one described so far.
We adopt a simple reference-counting scheme in which each \gls{rs} tracks, for every physical register, the number of consumers of the value it holds (\lstinline{consumer_count}) and reclaims the register only once all consumers have been served (\lstinline{consumed_by == consumer_count}).
An instruction can then start a new iteration as soon as the physical register allocated to the previous iteration has been reclaimed.
This enables instructions from different iterations to be simultaneously in flight across \glspl{rs} without violating false dependencies.
If a result is simultaneously consumed by multiple instructions, the \lstinline{consumed_by} counter needs to be incremented by the number of simultaneous consumers.
To this end, the result requests arriving at the \gls{rs} carry a bitmask (\lstinline{dest_mask}) identifying all the consumers requesting a result.
A popcount unit in each \lstinline{res_req_handler} module counts the number of consumers in the bitmask and increments the \lstinline{consumed_by} counter of the corresponding physical register accordingly.
During \gls{dcp}, the same mechanism is used to determine and record the total \lstinline{consumer_count} for each physical register.

As described in \Cref{sec:architecture}, the \gls{slp} architecture leverages the assumption that the dataflow graph remains static across iterations.
To satisfy this requirement, we restrict the approach to straight-line loop bodies, that is, loop bodies consisting of a single basic block \cite{aho2006}.
In such loops, any loop-carried define-use dependence through a fixed architectural register has distance one.
By definition, a value produced in iteration $i$ can reach a use in a later iteration only along a definition-clear path \cite{aho2006}, and in a straight-line loop, such a path cannot cross more than one loop boundary.
Consequently, the maximum loop-carried dependence distance \cite{allen2001} in our design is $n=1$%
\footnote{This does not mean \gls{slp} mode is restricted to source-level loops whose loop-carried dependences have distance $\leq 1$. This applies only to their assembly-level realization: for straight-line loop bodies, a dependence through a fixed architectural register cannot span more than one loop boundary. Source-level dependences of greater distance are still supported and are typically realized through different architectural registers.}.
In other words, an instruction issued in iteration $i$ can request operands produced either in the same iteration ($i$) or in the previous one ($i-1$).

To preserve true dependencies, each physical register is additionally tagged with the \lstinline{iteration} flag $i$ of the instruction that produced its value.
When an instruction issued in iteration $i$ requests an operand produced in the same iteration ($i$) or in the previous ($i-1$), it includes this information in the request.
At the producer side, the \lstinline{iteration} tag carried by the request is compared against that of the produced value, and the request is stalled until the two tags match.
In practice, storing the full iteration index $i$ is unnecessary; it is sufficient to store $i \mod (n + 1)$, where $n$ denotes the maximum loop-carried dependence distance among in-flight instructions.
As discussed, $n=1$ in our case, so the iteration index reduces to a 1-bit \lstinline{iteration} tag.

\subsection{Operand Delivery Network}
\label{sec:odn}

As mentioned in \Cref{sec:rs}, Schnizo selects at most one instruction per \gls{rs} at any time to request operands from the \gls{odn}.
In contrast, modern \gls{ooo} processors typically maintain more in-flight instructions waiting in the issue queues than can be issued simultaneously \cite{palacharla1997complexity, kessler1999}.
As a result, they usually implement both a wakeup network, to drive the issue-queue wakeup/select logic, and a separate operand network, comprising the \gls{prf} read path and bypass network, to retrieve and forward operand values to the selected instructions \cite{taylor2005}.
In Schnizo the selection logic is static, so a separate wakeup network is not required.
The \gls{odn} thus implements only the operand network, namely a mechanism to request and forward operands from producers to consumers.

The \gls{odn} handles two transactions: \textit{operand requests}, issued by consumers (\glspl{rs}) to request a value from a specific producer, and \textit{operand responses}, issued by producers to return that value.

As described in \Cref{sec:rs}, every \gls{rs} places a request on the \gls{odn} for each dynamic operand required by its currently selected instruction.
Each operand request carries two pieces of information: a tag identifying the producer's physical register holding the value, and a 1-bit \lstinline{iteration} tag, as described in \Cref{sec:rs}.

\begin{figure}[t!]
  \centering
  \includegraphics[width=\columnwidth]{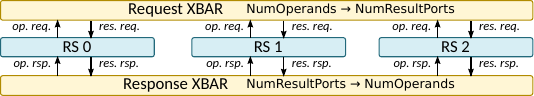}
  \caption{Architecture of the \gls{odn}.}
  \label{fig:odn}
\end{figure}

As shown in \Cref{fig:odn}, all operand requests are routed to the producers through the request \gls{xbar}.
The request \gls{xbar} provides full connectivity between the consumer \glspl{rs} and producer \glspl{rse}, allowing requests targeting different \glspl{rse} to be served in parallel.
When multiple requests target the same \gls{rse}, they may either request the same value or differ in the 1-bit \lstinline{iteration} tag.
In the latter case, the request that matches the \lstinline{iteration} of the currently available result is served.
In the former case, matching requests are coalesced into a single request, and the corresponding response is multicast to all requesters through the response \gls{xbar}.
To this end, matching requests are collected into a bitmask (\lstinline{dest_mask}), where each bit identifies an operand-request port.

Upon receiving a result request, the target \gls{rs} checks whether the requested value is available and whether the request matches its \lstinline{iteration} tag.
If both conditions hold, it drives the value together with the \lstinline{dest_mask} onto the response \gls{xbar}.

The response \gls{xbar} returns the values requested on the request \gls{xbar} back to the consumers that placed the requests.
It provides full crossbar connectivity between the producer (\lstinline{result}) ports and consumer (\lstinline{operand}) ports, allowing responses targeting different consumer ports to be delivered in parallel.
The response \gls{xbar} is multicast-capable: a value on a producer port can be delivered simultaneously to all consumer ports recorded in the \lstinline{dest_mask}.
Because packets on the response \gls{xbar} carry operand values of up to 64-bits, its complexity is higher than that of the request \gls{xbar}, whose payload consists only of the \lstinline{iteration} tag and producer \gls{rse}.
As a result, the number of read ports into each \gls{rs}'s \gls{prf}, which equals the number of producer ports on the response \gls{xbar}, has a significant impact on the complexity of the \gls{odn}.
We evaluate this tradeoff in \Cref{sec:scalability}.
On the consumer side, when an operand response arrives, the \gls{rs} captures the value in its operand buffers and updates the corresponding readiness information.

\subsection{Loop Controller and Register Renaming}
\label{sec:controller}

The remaining components of the \gls{slp} extension are the loop controller and the dispatch stage.

The loop controller manages processor state in the presence of \glspl{zol}.
From its perspective, the processor can operate in one of five execution modes: regular, \gls{zol}, \gls{dcp1}, \gls{dcp2} and \gls{dep}.
In regular mode, the processor executes instructions one by one, as in a conventional scalar core.
When a \gls{zol} instruction is fetched, the controller transitions to either \gls{zol} or \gls{dcp1} mode, depending on the value of the \lstinline{enable_slp} flag introduced in \Cref{sec:implementation}.
As described earlier, execution in the \gls{dcp} states remains in order, with the instruction frontend still responsible for dispatching instructions.
Relative to regular and \gls{zol} execution, the key differences are that 1) all \glspl{fu} are considered for dispatch, 2) the register-renaming stage is enabled, and 3) instructions are routed to the \glspl{rs}, where the associated bookkeeping information is updated.
When multiple \glspl{fu} of the same type are present, instructions are assigned to them in round-robin order to balance the load during superscalar execution.
Finally, if an unsupported instruction is encountered, as listed in \Cref{sec:limitations}, the core falls back to \gls{zol} without any performance penalty: no work is lost and no instructions are replayed.
The loop controller transitions to \gls{zol} state for the remaining instructions and loop iterations, while discarding the recorded \gls{dcp} state.

While \Cref{sec:architecture} described the \gls{dcp} phase under a simplified assumption, loop-carried dependencies require two \gls{dcp} passes over the loop body to fully resolve operand sources for all instructions.
Consider the dot product kernel in \Cref{fig:zol}, which contains a loop-carried dependency from instruction 5 to instruction 2 through register \lstinline{t0}.
When instruction 2 is dispatched during \gls{dcp1}, the \gls{rmt} entry for \lstinline{t0} is still empty since instruction 5 has not yet been encountered.
Without a mechanism to update the \gls{rse} holding instruction 2 after instruction 5 is dispatched in \gls{dcp1}, that is, once the producer of \lstinline{t0} has been identified and recorded in the \gls{rmt}, the \gls{rs} would incorrectly classify the operand as static, causing instruction 2 to use the wrong value in subsequent iterations.

To resolve this issue, we introduce a second \gls{dcp} iteration.
When the first loop iteration completes (\gls{dcp1}), the loop controller transitions to the \gls{dcp2} state and branches back to the first loop instruction to execute the second loop iteration of the original program.
In this phase, like in \gls{dcp1}, instructions are still issued one by one, in order, by the instruction frontend.
When instruction 2 is dispatched during \gls{dcp2}, the \gls{rmt} entry for \lstinline{t0} now points to the \gls{prf} entry allocated to instruction 5, allowing the correct producer to be recorded in the \gls{rse} holding instruction 2.
In addition, each \gls{rse} stores a 1-bit flag (\lstinline{is_from_prev_iter}) for every operand, which is set if the producer is recorded during \gls{dcp2} and cleared otherwise.
This flag indicates whether the operand is produced in the same iteration in which it is consumed or in the previous one.
During \gls{dep}, the \gls{rs} consults this flag when issuing operand requests on the \gls{odn} to drive the 1-bit \lstinline{iteration} tag described in \Cref{sec:rs}.
Operands that are still not mapped in the \gls{rmt} during \gls{dcp2} are guaranteed to be static and are therefore written by the \gls{rs} into its \gls{cm}.

Similarly, the \gls{dcp2} phase is also used to resolve the \lstinline{do_writeback} flag, which determines whether an instruction updates the architectural register state in the final loop iteration, as described in \Cref{sec:rs}.
When an instruction is dispatched during \gls{dcp1}, it is not yet possible to determine whether it is the last instruction in program order to write its destination register.
To address this, the \gls{rmt} state is frozen at the end of \gls{dcp1}.
When the instruction is revisited during \gls{dcp2}, the \gls{rmt} is consulted to determine whether the last producer recorded for its destination register is the instruction itself; if so, the \lstinline{do_writeback} flag is set.

Upon completion of \gls{dcp2}, the loop controller transitions to \gls{dep}, during which the instruction frontend is disabled and the \glspl{rs} operate independently.
The loop controller then monitors the progress of the \glspl{rs} and determines loop completion by aggregating their individual completion states.

\subsection{Limitations}
\label{sec:limitations}

\begin{figure*}[t!]
  \centering
  \begin{minipage}{0.34\textwidth}
    \begin{subfigure}{\textwidth}
      \includegraphics[width=\textwidth]{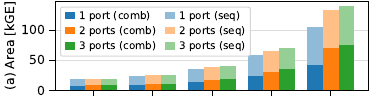}
      \vspace{-1em}
      \phantomcaption
      \label{fig:rs_area}
    \end{subfigure}
    \begin{subfigure}{\textwidth}
      \includegraphics[width=\textwidth]{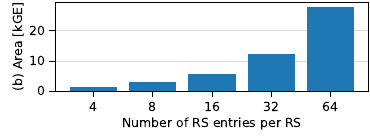}
      \vspace{-1em}
      \phantomcaption
      \label{fig:req_xbar_area_slots}
    \end{subfigure}
  \end{minipage}%
  \hfill
  \begin{minipage}{0.28\textwidth}
    \begin{subfigure}{\textwidth}
      \includegraphics[width=\textwidth]{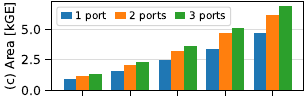}
      \vspace{-1em}
      \phantomcaption
      \label{fig:req_xbar_area}
    \end{subfigure}
    \begin{subfigure}{\textwidth}
      \includegraphics[width=\textwidth]{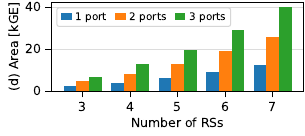}
      \vspace{-1em}
      \phantomcaption
      \label{fig:rsp_xbar_area}
    \end{subfigure}
  \end{minipage}%
  \hfill
  \begin{minipage}{0.34\textwidth}
    \begin{subfigure}{\textwidth}
      \includegraphics[width=\textwidth]{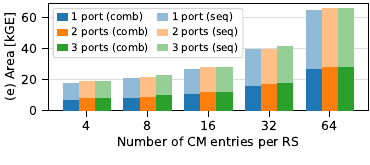}
      \vspace{-1em}
      \phantomcaption
      \label{fig:rs_constants_area}
    \end{subfigure}
  \end{minipage}
  \caption{
    Area of the \gls{rs} as a function of the number of \glspl{rse} (\subref{fig:rs_area}) and \glspl{cme} (\subref{fig:rs_constants_area}); area of the request \gls{xbar} as a function of the number of \glspl{rse} (\subref{fig:req_xbar_area_slots}) and \glspl{rs} (\subref{fig:req_xbar_area}); and area of the response \glspl{xbar} as a function of the number of \glspl{rs} (\subref{fig:rsp_xbar_area}).}
  \label{fig:area}
\end{figure*}

As motivated in \Cref{sec:introduction}, the \gls{slp} architecture seeks to balance performance and flexibility against hardware cost and efficiency.
It does so by specializing to a class of regular programs, thereby enabling a simpler and more efficient design.

As described in \Cref{sec:rs}, the main limitation of the current \gls{slp} design is the lack of support for control flow within the loop body, as the dataflow graph must remain static across iterations.
As a result, conditional branches (\lstinline{beq}, \lstinline{bne}, and similar instructions) and indirect jumps (\lstinline{jr}, \lstinline{jalr}) are disallowed.
These instructions are detected at runtime during \gls{dcp} and handled gracefully by falling back to regular \gls{zol} execution.
This limitation could further be partially mitigated through predication \cite{allen1983, mahlke1992}.
By converting control dependences into data dependences, predication preserves a straight-line loop body and a static dataflow graph across iterations, making it conceptually compatible with the \gls{slp} execution model.
Another possible mitigation is to extend the \gls{zol} logic to support nested loops, which prior work has shown can be achieved at low cost \cite{colagrande2025gemm}.
For brevity, we leave the exploration of such techniques to future work, where their hardware/software implications and quantitative impact can be studied in depth.
Unconditional jumps (\lstinline{j}, \lstinline{jal}) are supported in principle, but only if the resulting inlined loop body still fits within the available \gls{rs} capacity.

Several resource limitations constrain the applicability of the \gls{slp} approach to generic workloads.
These include running out of \glspl{rse} or \glspl{cme} in any target \gls{fu}, as well as exceeding the maximum supported consumer count for a result.
Again, all such conditions are detected at runtime during \gls{dcp} and handled gracefully by falling back to regular \gls{zol} execution.
These limitations could also be mitigated through compiler transformations such as loop fission (or distribution) \cite{allen2001, kennedy1990, banerjee1993}, which partitions a large loop body into multiple smaller loops that fit within the available \gls{slp} resources.
This trades some of the benefits of a single larger loop body for lower per-loop resource requirements.
Evaluating the effectiveness of these techniques for \gls{slp} is left for future work.

Furthermore, although this is not an intrinsic limitation of the \gls{slp} architecture, the current Schnizo implementation does not enforce any ordering guarantees between memory operations mapped to different \glspl{lsu}.
In the presence of memory dependencies, e.g. a load following (in program order) a store to the same address, this can result in incorrect single-thread behaviour.
To prevent this, by default Schnizo's dispatcher assigns all memory instructions to a single LSU, even if multiple LSUs are present in the hardware, serializing all memory operations and therefore guaranteeing their in-order execution.
We note that this constraint can result in a significant performance limitation for low arithmetic intensity kernels, such as the AXPY and DOT kernels adopted in our evaluation.
However, this constraint is only required for correctness in the presence of dependent memory operations.
It can be safely relaxed when all memory operations in a loop can be guaranteed to be independent.
This is the case for all but the EXP kernel adopted in our evaluation.
To this end, we provide an additional bit in the \lstinline{zol} instruction (\lstinline{relax_mem.}), shown in \Cref{fig:zol_instruction}, which allows the programmer or compiler to relax this constraint on a per-loop basis, when all memory operations in a loop can be proven independent, e.g. through pointer-alias analysis \cite{hind2001,banerjee1997}.
Accordingly, in our evaluation we set the \lstinline{relax_mem.} bit to 1 for all but the EXP kernel.

Finally, while precise exceptions are supported in regular, \gls{zol} and \gls{dcp} modes, they are not supported in \gls{dep}, as \gls{dep} does not enforce in-order commit.
Support for exceptions in \gls{dep} could nevertheless be added by treating \gls{dep} as a speculative region and rolling back to the pre-\gls{dep} state in the event of an exception.
Since the architectural \gls{rf} remains unchanged until the final loop iteration, the \gls{rf} state can be restored without additional hardware, by deferring writeback to the architectural \gls{rf} until loop termination.
For exceptions that occur after stores have been committed to the memory hierarchy, however, \gls{rf} rollback alone cannot restore the full pre-loop architectural state.
Restart-based recovery remains safe in this case only if the loop writes to output-only locations disjoint from its inputs (i.e., the loop is idempotent \cite{dekruijf2011}): re-execution will not be affected by the non-restored output values, and these values will be overwritten.
If, on the other hand, stores overwrite loop inputs, update shared state, or otherwise create externally visible side effects that cannot be undone, \gls{rf} rollback alone is insufficient.
Supporting precise \gls{dep} exceptions for such loops would require additional mechanisms, such as store buffering, which we leave to future work.
We further note that, in the intended deployment context of \glspl{slp}, namely high-throughput programmable accelerators targeting regular loop kernels, strict precise-exception support is often relaxed in practice.
In such systems, exceptions are commonly employed as mechanisms for debugging, error reporting, or kernel failure detection, rather than as a general-purpose precise trap-and-resume control-flow mechanism comparable to CPU exceptions.
\glspl{gpu} for instance often report exceptions imprecisely \cite{menon2012igpu,tanasic2017,nvidia_cuda_gdb_errors}, omit hardware trap support for certain exception classes, such as floating-point exceptions, \cite{li2023gpufpx} or support precise reporting through optional debug modes or single-stepping mechanisms, which trade performance for precision \cite{amd_rocgdb_memory,amd_rocgdb_alu}.
In Schnizo, regular or \gls{zol} execution provides such a precise fallback mode, while \gls{dep} should be understood as the high-throughput mode in which exceptions are imprecise unless restartability assumptions are met.
\section{Results}
\label{sec:results}

For area evaluation, we implement the \gls{rs}, the request and response \glspl{xbar}, and the full Schnizo core in GlobalFoundries' \SI{12}{\nano\meter} LP+ technology using Fusion Compiler 2023.12, with a \SI{1}{\giga\hertz} frequency target under worst-case conditions (SS, \SI{125}{\celsius}, \SI{0.72}{\volt}).
For performance evaluation, we integrate Schnizo into an L1 cluster equipped with a local L1 \gls{spm} and I\$, derived from the Snitch cluster \cite{zaruba2021snitch}.
Performance is evaluated through cycle-accurate RTL simulation of the Schnizo cluster in QuestaSim 2023.4 at the target frequency.
All benchmark codes are implemented in bare-metal C++ and inline assembly, compiled with a custom C++ runtime and compiler toolchain for Snitch based on LLVM 15 at the highest optimization level (-O3).
All codes are further manually optimized, through a combination of software pipelining and 4$\times$ loop unrolling, to hide the 4-cycle \gls{fpu} and L1 memory latencies.
For transparency and reproducibility, all codes are open source\footnotemark.

\subsection{Scalability}
\label{sec:scalability}

\footnotetext{
\ifdefined\blindreview
    https://hidden-for-double-blind-review.com
\else
    Refer to commit \href{https://github.com/colluca/schnizo/tree/dd1bb60472fc27c599079eb10e977cb662e2f143}{dd1bb60472fc27c599079eb10e977cb662e2f143}.
\fi
}

We evaluate the scalability of Schnizo by varying the number of \glspl{rs}, \glspl{rse}, \glspl{cme} and result ports on the \gls{odn}.
The number of \glspl{rs} determines the degree of parallelism that can be extracted across \glspl{fu}, while the number of \glspl{rse} determines the maximum loop-body size that can be supported in superscalar execution, and more generally the flexibility of \gls{slp} mode.
Similarly, the number of \glspl{cme} determines the maximum number of static operands supported in \gls{slp} mode.
The number of result ports determines how many operands an \gls{rs} can deliver in parallel, which can affect performance for workloads with imbalanced producer-consumer patterns.

\Cref{fig:rs_area} shows the area of the \gls{rs} as a function of the number of \glspl{rse}, for a fixed number of \glspl{cme} (NumCme=4).
We further vary the number of result ports.
The results show that area scales near-linearly with the number of \glspl{rse} ($R^2 > \ResultRsAreaMinRsq$ in all cases).
Specifically, total area increases by \SI{\ResultRsAreaOnePortTotSlopeKgePerRse}{\kilo\ge}, \SI{\ResultRsAreaTwoPortTotSlopeKgePerRse}{\kilo\ge} and \SI{\ResultRsAreaThreePortTotSlopeKgePerRse}{\kilo\ge} per additional \gls{rse} for 1-, 2- and 3-port \glspl{rs}, respectively.
Sequential and combinational logic contribute similarly: sequential logic is dominated by the flip-flop-heavy \gls{rs} storage, whereas combinational logic is dominated primarily by operand multiplexing.
For a fixed number of \glspl{rse}, \gls{rs} area also scales with the number of \glspl{cme}, as shown in \Cref{fig:rs_constants_area} (NumRse=4).
Again, the trend is near-linear ($R^2 > \ResultRsConstantsMinRsq$), with an increase of \SI{\ResultRsConstantsOnePortTotSlopeKgePerConstant}{\kilo\ge}, \SI{\ResultRsConstantsTwoPortTotSlopeKgePerConstant}{\kilo\ge} and \SI{\ResultRsConstantsThreePortTotSlopeKgePerConstant}{\kilo\ge} per additional \gls{cme} for 1-, 2- and 3-port \glspl{rs}, respectively, as the constant storage is independent of the number of result ports.

\Cref{fig:req_xbar_area,fig:rsp_xbar_area} show the area of the request and response \glspl{xbar}, respectively, as a function of the number of \glspl{rs}.
We further vary the number of result ports, as described above.
Both \glspl{xbar} are purely combinational and scale near-linearly with the number of \glspl{rs}: the request \gls{xbar} at \SI{\ResultReqXbarOnePortSlopeKgePerRs}, \SI{\ResultReqXbarTwoPortSlopeKgePerRs}, and \SI{\ResultReqXbarThreePortSlopeKgePerRs}{\kilo\ge}/\gls{rs} ($R^2 > \ResultRspXbarMinRsq$) for 1, 2, and 3 result ports, respectively; and the response \gls{xbar} at \SI{\ResultRspXbarOnePortSlopeKgePerRs}, \SI{\ResultRspXbarTwoPortSlopeKgePerRs}, and \SI{\ResultRspXbarThreePortSlopeKgePerRs}{\kilo\ge}/\gls{rs} ($R^2 > \ResultRspXbarMinRsq$).
As we can see, increasing the number of result ports on the request and response \glspl{xbar} itself leads to a near-linear increase in the cost per \gls{rs}, at \SI{\ResultReqXbarCostPerPortKgePerRs}{\kilo\ge}/\gls{rs} ($R^2 = \ResultReqXbarCostPerPortRsq$) and \SI{\ResultRspXbarCostPerPortKgePerRs}{\kilo\ge}/\gls{rs} ($R^2 = \ResultRspXbarCostPerPortRsq$) per additional result port on the request and response \glspl{xbar}, respectively, reducing the scalability of the architecture.
In \Cref{sec:tradeoffs}, we explore the performance implications of reducing the number of result ports.
Due to the coalescing logic described in \Cref{sec:odn}, the request \gls{xbar} area also grows with the number of \glspl{rse} per \gls{rs}, as shown in \Cref{fig:req_xbar_area_slots} (NumResultPorts=3,NumRs=3). Specifically, we observe a near-linear growth ($R^2 > \ResultReqXbarSlotsThreePortRsq$), at \SI{\ResultReqXbarSlotsThreePortSlopeKgePerRse}{\kilo\ge}/\gls{rse}.

\begin{table}[t!]
  \centering
  \caption{
    Scalar and superscalar core configurations with varying number of \glspl{fu} (F), \gls{rs} entries (R) and \gls{cm} entries (C).
  }
  \label{tab:configs}
  \centering
  \begin{threeparttable}
    {
    \setlength{\tabcolsep}{3pt}
    \begin{tabular}{l rrr rrr rrr rrrr r r}
      \hline
      \multirow{2}{*}{Config.} & \multicolumn{3}{c}{ALU} & \multicolumn{3}{c}{LSU} & \multicolumn{3}{c}{FPU} & \multicolumn{3}{c}{Tot.} & \multirow{2}{*}{\makecell[r]{\textsuperscript{\S}Area\\{[kGE]}}} & \multirow{2}{*}{\makecell[r]{\textsuperscript{\dag}Area\\Incr.}} \\
      \cmidrule(lr){2-4} \cmidrule(lr){5-7} \cmidrule(lr){8-10} \cmidrule(lr){11-13}
      & F & R & C & F & R & C & F & R & C & F & R & C & & \\
      \hline
      Snitch   & 1 & -   & -  & 1 & -  & -  & 0 & -  & -  & 2  & -   & -   & \ResultSnitchAreaKge                 & \ResultSnitchAreaRatio$\times$         \\
      Schnizo  & 1 & -   & -  & 1 & -  & -  & 0 & -  & -  & 2  & -   & -   & \ResultScalarAreaKge                 & \ResultScalarAreaRatio$\times$         \\
      Schn.-FP & 1 & -   & -  & 1 & -  & -  & 1 & -  & -  & 3  & -   & -   & \ResultScalarMulFpuAreaKge           & \ResultScalarMulFpuAreaRatio$\times$   \\
      \hline
      Small  & 1 & 1   & 1  & 1 & 1  & 1  & 1 & 1  & 1  & 3  & 3   & 3   & \ResultSuperscalarSAreaKge           & \ResultSuperscalarSAreaRatio$\times$   \\
      LA     & 3 & 4   & 4  & 3 & 4  & 4  & 1 & 4  & 4  & 7  & 28  & 28  & \ResultSuperscalarLaAreaKge          & \ResultSuperscalarLaAreaRatio$\times$  \\
      MC     & 3 & 32  & 16 & 1 & -  & -  & 2 & 32 & 16 & 6  & 160 & 80  & \ResultSuperscalarMcAreaKge          & \ResultSuperscalarMcAreaRatio$\times$  \\
      TR     & 2 & 32  & 16 & 1 & 32 & 64 & 2 & 32 & 16 & 5  & 160 & 128 & \ResultSuperscalarTrAreaKge          & \ResultSuperscalarTrAreaRatio$\times$  \\
      GP-S   & 1 & 128 & 32 & 1 & 32 & 64 & 1 & 64 & 32 & 3  & 224 & 128 & \ResultSuperscalarGpMAreaKge         & \ResultSuperscalarGpMAreaRatio$\times$ \\
      GP-L   & 3 & 32  & 16 & 3 & 32 & 64 & 1 & 64 & 32 & 7  & 256 & 272 & \ResultSuperscalarGpLAreaKge         & \ResultSuperscalarGpLAreaRatio$\times$ \\
      \hline
    \end{tabular}
    }
    \begin{tablenotes}
      \footnotesize
      \item[\S] Does not include FP DIV/SQRT unit.
      \item[\dag] Relative to the \schnizo{fp} configuration.
    \end{tablenotes}
  \end{threeparttable}
\end{table}

To evaluate the overall area of Schnizo, we implement the different configurations summarized in \Cref{tab:configs}.
As shown, both Snitch and the scalar Schnizo configuration fit within a modest footprint of at most $\leq\SI{\ResultSnitchAreaKge}{\kilo\ge}$, making them suitable for integration in both resource-constrained and massively-parallel \glspl{soc}.
The scalar \schnizo{fp} configuration further integrates Snitch's 64-bit \gls{simd}-capable \gls{fpu} as an additional \gls{fu}.
This configuration is equivalent to the Snitch core complex shown in \Cref{fig:snitch}, while saving on the \gls{zolb} and \gls{ssr}-dedicated area.
Hence, we use \schnizo{fp} as a baseline for the superscalar configurations.

We evaluate multiple superscalar configurations, tuned for different kernel groups, as defined in \Cref{tab:kernels}: linear algebra (LA), Monte Carlo (MC), and transcendental function (TR) kernels.
We further evaluate the smallest theoretically-possible configuration (Small) and two general-purpose (GP) configurations balanced to meet the requirements of all microkernels: a smaller one (GP-Small) with 3 \glspl{fu}, and a larger one (GP-Large) with 7 \glspl{fu}.
We order the configurations in \Cref{tab:configs} by total number of \glspl{rse} and \glspl{cme}.
As shown, the Small configuration, with a single \gls{rse} and \gls{cme} per \gls{fu}, introduces only a \ResultSuperscalarSOverheadPct\% overhead relative to \schnizo{fp}, while theoretically tripling \gls{ipc}, thanks to the low complexity of Schnizo's superscalar scheduling logic.
In practice, however, larger \gls{rse} and \gls{cme} counts are required to support realistic loop bodies, as discussed in \Cref{sec:limitations}.
The smallest useful configuration (LA), providing up to 28 \glspl{rse} and \glspl{cme}, incurs an area overhead of \ResultSuperscalarLaOverheadPct\% relative to \schnizo{fp}.

More broadly, total core area increases monotonically with the number of \glspl{rse} and \glspl{cme}, consistent with the trends observed in \Cref{fig:area}, indicating that area growth in Schnizo is driven primarily by total \gls{rs} storage rather than by the number of \glspl{rs}.
This exposes a direct flexibility-cost tradeoff in Schnizo: supporting larger, more complex, or more heavily unrolled loop bodies requires provisioning additional \glspl{rse} and \glspl{cme}, and the cost of this increased architectural flexibility grows almost linearly with \gls{rs} storage, as observed above.
By contrast, increasing the number of \glspl{rs}, and thus the degree of parallelism exposed by the architecture, incurs only a modest marginal cost, as Schnizo's scheduling logic scales gracefully with issue width.

\subsection{Performance}
\label{sec:performance}

\begin{table}[t]
  \caption{
    Microkernel loop body instruction mix.
    Reported instruction counts already account for unrolling.
  }
  \label{tab:kernels}
  \centering
  \begin{threeparttable}
    {
    \setlength{\tabcolsep}{3pt}
    \begin{tabular}{l c c rrrrrrr}
      \hline
      \multirow{2}{*}{Kernel} & \multirow{2}{*}{Group} & \multirow{2}{*}{\makecell[c]{Unr.\\Fac.}} & \multicolumn{4}{c}{\#Instructions} & \multirow{2}{*}{\makecell[c]{\#Co-\\nst.}} & \multicolumn{2}{c}{Ideal \gls{ipc}} \\
      \cmidrule(lr){4-7} \cmidrule(lr){9-10}
      & & & ALU & FPU & LSU & Tot. & & GP-L & Spec. \\
      \hline
      axpy              & LA & \textsuperscript{\dag}1$\times$ & \ResultSzAxpyAluInsns       & \ResultSzAxpyFpuInsns       & \ResultSzAxpyLsuInsns       & \ResultSzAxpyTotalInsns       &  7 & \ResultSzAxpyIdealGpIpc       & \ResultSzAxpyIdealIpc       \\
      dot               & LA & 4$\times$                       & \ResultSzDotAluInsns        & \ResultSzDotFpuInsns        & \ResultSzDotLsuInsns        & \ResultSzDotTotalInsns        & 10 & \ResultSzDotIdealGpIpc        & \ResultSzDotIdealIpc        \\
      exp               & TR & 4$\times$                       & \ResultExpAluInsns          & \ResultExpFpuInsns          & \ResultExpLsuInsns          & \ResultExpTotalInsns          & 86 & \ResultExpIdealGpIpc          & \ResultExpIdealIpc          \\
      log               & TR & 4$\times$                       & \ResultLogAluInsns          & \ResultLogFpuInsns          & \ResultLogLsuInsns          & \ResultLogTotalInsns          & 62 & \ResultLogIdealGpIpc          & \ResultLogIdealIpc          \\
      pi\_lcg           & MC & 4$\times$                       & \ResultPiLcgAluInsns        & \ResultPiLcgFpuInsns        & \ResultPiLcgLsuInsns        & \ResultPiLcgTotalInsns        & 28 & \ResultPiLcgIdealGpIpc        & \ResultPiLcgIdealIpc        \\
      pi\_xoshiro       & MC & 4$\times$                       & \ResultPiXoshiropAluInsns   & \ResultPiXoshiropFpuInsns   & \ResultPiXoshiropLsuInsns   & \ResultPiXoshiropTotalInsns   & 36 & \ResultPiXoshiropIdealGpIpc   & \ResultPiXoshiropIdealIpc   \\
      poly\_lcg         & MC & 4$\times$                       & \ResultPolyLcgAluInsns      & \ResultPolyLcgFpuInsns      & \ResultPolyLcgLsuInsns      & \ResultPolyLcgTotalInsns      & 32 & \ResultPolyLcgIdealGpIpc      & \ResultPolyLcgIdealIpc      \\
      poly\_xoshiro     & MC & 4$\times$                       & \ResultPolyXoshiropAluInsns & \ResultPolyXoshiropFpuInsns & \ResultPolyXoshiropLsuInsns & \ResultPolyXoshiropTotalInsns & 40 & \ResultPolyXoshiropIdealGpIpc & \ResultPolyXoshiropIdealIpc \\
      \hline
    \end{tabular}
    }
    \begin{tablenotes}
      \footnotesize
      \item[\dag]4$\times$ in scalar Snitch and Schnizo configurations.
    \end{tablenotes}
  \end{threeparttable}
\end{table}

We evaluate Schnizo's performance on a set of microkernels extracted from, or representative of, a range of \gls{ml} and \gls{hpc} applications, ported from Snitch \cite{zaruba2021snitch,colagrande2025copift}.
We adopt these kernels for a detailed design-space analysis, as they exhibit diverse instruction mixes and dependency patterns.
\Cref{tab:kernels} reports the instruction mix of each microkernel, broken down by \gls{fu} type, together with its ideal \gls{ipc}, defined as:
\begin{equation}
  \text{IPC}_\text{ideal} = \frac{\#\text{Insn}_\text{tot}}{\max_i \frac{\#\text{Insn}_i}{\#\text{FU}_i}}, i \in \{\text{ALU, FPU, LSU}\}
\end{equation}
The denominator is the number of cycles required by the most heavily utilized \gls{fu}, under the idealized assumption that each \gls{fu} can sustain a throughput of one instruction per cycle and that data dependencies, scheduling effects, and other pipeline bottlenecks do not limit execution.
This metric therefore captures the maximum \gls{ilp} that a workload exhibits on an architecture with a given \gls{fu} mix.
If a workload is perfectly balanced across \glspl{fu}, that is, if $\frac{\#\text{Insn}_i}{\#\text{FU}_i}$ is equal for all $i$, then $\text{IPC}_\text{ideal} = \#\text{FU}_\text{tot}$, where $\#\text{FU}_\text{tot}$ denotes the total number of \glspl{fu}, i.e., the architecture's degree of parallelism.
For each workload, we compare the \gls{ipc} measured on Schnizo against this ideal \gls{ipc}, which provides an upper bound on the performance attainable by any concrete architecture with the same \gls{fu} mix.

To demonstrate Schnizo's wider applicability to \gls{ml} kernels, we further implement 11 workloads from the NonGEMM benchmark suite \cite{karami2025nongemm}, including all operators from the most time-consuming non-GEMM operator groups (excluding memory operators) listed by the authors in Table I, i.e., all normalization, element-wise arithmetic, activation, and logit computation operators.

\begin{figure}[t!]
  \centering
  \includegraphics[width=\columnwidth]{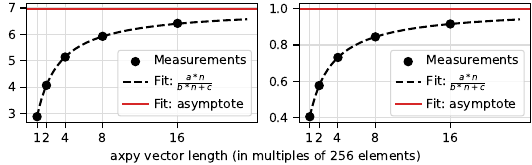}
  \caption{\gls{ipc} (left) and \gls{fpu} utilization (right) of the AXPY kernel on \schnizo{gp-l}, for varying problem sizes.}
  \label{fig:axpy_perf}
\end{figure}

We first evaluate the performance impact of the \gls{dcp} phases by benchmarking the AXPY kernel on \schnizo{gp-l} across varying problem sizes, as shown in \Cref{fig:axpy_perf}.
For smaller problem sizes, the sequential \gls{dcp} phases have a more pronounced impact, as they account for a larger fraction of total execution.
Fitting Amdahl's-law-derived curves \cite{amdahl1967} to the empirical data shows that, as this sequential fraction approaches zero, Schnizo's performance asymptotically approaches the ideal \gls{ipc} (7) and \gls{fpu} utilization (1) achieved in \gls{dep}.
We further note that this effect is theoretically independent of the loop body size: the fitted curve parameters $a$, $b$, and $c$ do not depend on the number of loop body instructions, as proven in \Cref{sec:appendix-slp-proof}.

In \Cref{fig:perf} we extend the analysis to all microkernels, reporting both the performance achieved at the largest problem size that fits within a \SI{128}{\kibi\byte} L1 \gls{spm} and the asymptotic performance.
With exception of the kernels involving element-wise divisions (DIV, GELU and SILU), which are bottlenecked by the multi-cycle latency of the FP division unit, almost all other microkernels achieve a significant fraction of their ideal \glspl{ipc}.
Across these kernels, Schnizo achieves a geomean \ResultGeomeanFractionOfIdealIpcExclDiv\% fraction of the ideal \glspl{ipc} and \ResultGeomeanFpuUtilExclDiv\% \gls{fpu} utilization at the maximum problem size, and a \ResultAsymptoticGeomeanFractionOfIdealIpcExclDiv\% fraction and \ResultAsymptoticGeomeanFpuUtilExclDiv\% utilization asymptotically, demonstrating Schnizo's effectiveness in extracting parallelism from real workloads.

To measure the impact on real-world \gls{ml} workloads, we profile \ResultNumNonGemmLayers\ non-GEMM layers extracted from 16 widely-adopted \gls{ml} models using the NonGEMM framework \cite{karami2025nongemm}.
All layers are mapped to our non-GEMM kernels, with parameters set to the corresponding layer-specific problem sizes.
\Cref{fig:non_gemm_perf} reports the speedup achieved on \schnizo{gp-l} through \gls{slp} execution, on the per-model cumulative runtime of all profiled layers.

\subsection{Tradeoffs}
\label{sec:tradeoffs}

\begin{figure}[t!]
  \centering
  \begin{subfigure}[t]{\columnwidth}
    \centering
    \includegraphics[width=\columnwidth]{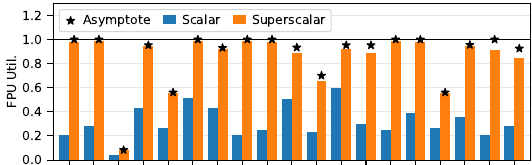}
  \end{subfigure}
  \begin{subfigure}[t]{\columnwidth}
    \centering
    \includegraphics[width=\columnwidth]{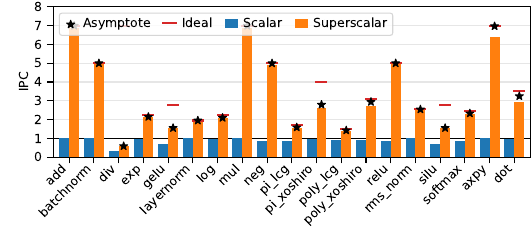}
  \end{subfigure}
  \caption{Scalar vs.\ superscalar performance across microkernels on the \schnizo{gp-l} configuration.}
  \label{fig:perf}
\end{figure}

\begin{figure*}[t!]
  \centering
  \begin{minipage}[t]{0.325\textwidth}
    \centering
    \includegraphics[width=\textwidth]{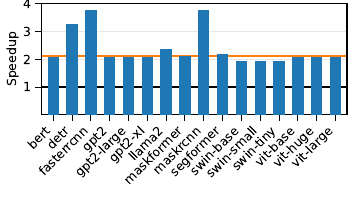}
    \caption{Speedup of non-GEMM operators across 16 \gls{ml} models (cumulative speedup in orange).}
    \label{fig:non_gemm_perf}
  \end{minipage}
  \hfill
  \begin{minipage}[t]{0.265\textwidth}
    \centering
    \includegraphics[width=\textwidth]{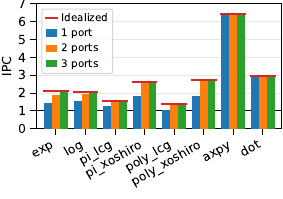}
    \caption{Performance for varying numbers of result ports per \gls{rs}.}
    \label{fig:rsp_ports_tradeoff}
  \end{minipage}
  \hfill
  \begin{minipage}[t]{0.385\textwidth}
    \centering
    \includegraphics[width=\textwidth]{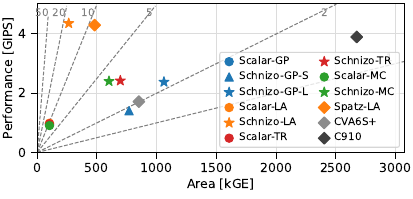}
    \caption{Area-performance comparison. Area-efficiency contour lines are represented in MIPS/kGE.}
    \label{fig:area_perf_tradeoff}
  \end{minipage}
\end{figure*}

The previous evaluation was conducted on the 7-\gls{fu} \schnizo{gp-l} configuration, assuming an idealized \gls{odn} with one result port per \gls{rse}, in order to stress the throughput of the architecture.
However, as shown in \Cref{sec:scalability}, the number of result ports has a significant impact on area.
In \Cref{fig:rsp_ports_tradeoff}, we therefore study the performance tradeoff as a function of the number of result ports per \gls{rs}.
The red lines indicate the performance levels reported in \Cref{fig:perf}.
As shown, three result ports per \gls{rs} are sufficient to reach \ResultGeomeanFractionOfFcIpcWithThreePorts\% of the geomean \gls{ipc} attained with the fully-connected \gls{odn}, while this fraction drops to \ResultGeomeanFractionOfFcIpcWithTwoPorts\% and \ResultGeomeanFractionOfFcIpcWithOnePort\% for two and one result ports, respectively.
Notably, AXPY and DOT are unaffected by the reduction in result ports because their balanced producer-consumer patterns never generate simultaneous requests for different results from the same producer \gls{rs}.

In \Cref{sec:scalability}, we observed that flexibility incurs a significant cost in area, and in \Cref{sec:performance} we evaluated performance on a flexible superscalar configuration that satisfies the \gls{rse} and \gls{cme} requirements of all microkernels.
We now examine Schnizo's area-performance tradeoff.
We evaluate the performance of the microkernel groups defined in \Cref{tab:kernels} on the respective specialized Schnizo configurations%
\footnote{This specialization limits the applicability of each configuration to other microkernel groups only as a result of resource limitations, as discussed in \Cref{sec:limitations}.
For example, the \lstinline{exp} kernel cannot be trivially mapped onto \schnizo{la} due to insufficient \glspl{rse}.
This could be enabled through transformations, such as loop fission, that we discuss in \Cref{sec:limitations} but do not explore in this work.
This section focuses specifically on this area-flexibility tradeoff, using the selected microkernels for illustrative purposes only.}
presented in \Cref{tab:configs}.
All configurations feature less than or equal to 7 \glspl{fu} as in \schnizo{gp-l}, and reallocate them to better match the instruction mix of each group, increasing geomean ideal \gls{ipc} across all microkernels by \ResultGeomeanIdealIpcIncrOverGpl\% to \ResultGeomeanIdealIpc\ (\Cref{tab:kernels}).
\Cref{fig:area_perf_tradeoff} reports geomean performance across the microkernels in each group.
The scalar datapoints all correspond to the same \schnizo{fp} configuration and therefore have identical area, but are evaluated separately on the different microkernel groups; the resulting differences are minimal, as all microkernels achieve \glspl{ipc} close to 1 (\Cref{fig:perf}).

Notably, \schnizo{la} achieves an area efficiency of \ResultLaGeomeanAreaEfficiency\,MIPS/kGE, \ResultLaGeomeanAreaEfficiencyImpr\% higher than its scalar counterpart.
All other Schnizo configurations fall short of their scalar counterparts, albeit delivering geomean speedups of up to \ResultMaxGeomeanSpeedupExclLa$\times$ across configurations.
The core-level area-efficiency metric is conservative for the many-core setting targeted by \glspl{slp}.
At cluster or chip level, improvements in \gls{ipc} translate almost directly into higher throughput, whereas increases in \gls{pe} area are partially amortized by fixed non-core overheads, such as L1 memories and interconnects.
To account for this effect, we add a fixed \SI{223}{\kilo\ge} overhead per core, corresponding to the per-core share of an 8-core Snitch cluster's L1 memory and interconnect area \cite{zaruba2021snitch}.
Under this model, Schnizo-LA's area-efficiency advantage over the scalar baseline increases from \ResultLaGeomeanAreaEfficiencyImpr\% to \ResultLaGeomeanAreaEfficiencyImprCluster\%.
 
To put the Schnizo-LA datapoint in context, we compare it against a recent open-source vector processor, Spatz, a state-of-the-art area-efficient architecture for vectorizable kernels.
The Spatz-LA datapoint reports the area of a 4-lane double-precision Spatz configuration and its geomean performance on AXPY and DOT, taken from TROOP \cite{purayil2025troop} and normalized by the number of lanes.
Performance numbers were calculated as the equivalent number of scalar instructions per cycle required to achieve the reported \gls{fpu} utilizations on a scalar RISC-V core, multiplied by Spatz' \SI{1}{\giga\hertz} clock frequency.
While Spatz achieves a comparable performance per lane to \schnizo{la}, its area efficiency is \ResultSpatzLaGeomeanAreaEfficiencyDecrease\% lower, owing to the vector \gls{rf} alone accounting for \ResultSpatzVrfAreaToSchnizoLaAreaRatio\% of \schnizo{la}'s total area.

Finally, the figure also compares Schnizo against open-source \gls{soa} \gls{ooo} processors.
Unfortunately, to the best of our knowledge, the only modern superscalar processors for which public core-only area data, excluding caches, is available are CVA6S+ \cite{fu2025} and C910 \cite{c910}.
To enable a benchmark-agnostic comparison, we plot each processor's theoretical peak performance, defined as the product of its issue width and reported clock frequency.
Even under this unfavorable assumption, the specialized Schnizo configurations still achieve higher performance and area efficiency, primarily due to CVA6S+'s and C910's comparably lower issue widths (2 and 3, respectively).
This highlights the different design point: conventional \gls{ooo} cores incur significant area overheads to retain general-purpose flexibility even at modest issue widths, whereas Schnizo specializes for regular loops to expose wider loop-mode issue with lower-complexity hardware.
We note, however, that this is not intended as a direct or fully fair comparison in terms of area and area efficiency, since these processors target fully-flexible general-purpose computing, whereas Schnizo is tailored to accelerator-oriented workloads.
Accordingly, we expect a significant portion of their area to be devoted to features, such as branch prediction, that are not essential for accelerator-oriented workloads \cite{iliakis2022}.
A fair comparison against a stripped-down \gls{ooo} core will be developed in future work.
Nevertheless, the comparison suggests that performance competitive with conventional \gls{ooo} architectures can be achieved at a reduced area footprint, and that \glspl{slp} represent an effective solution in this direction.

\section{Related Work}
\label{related_work}

\emph{Loop-oriented specializations.} %
In \Cref{sec:background} we discussed how our work builds on prior loop-oriented processor specializations, including hardware loops \cite{bajwa1997, uh1999, tsao2003, talla2003, kavvadias2008, kavvadias2010, wei2010, vadivel2017, gautschi2017pulp, colagrande2025gemm}, \acrfullpl{zolb} \cite{bajwa1997, uh1999, tsao2003, vadivel2017, wei2010, colagrande2025gemm, zaruba2021snitch}, Snitch's pseudo-dual-issue execution model \cite{zaruba2021snitch}, and its software-assisted extensions COPIFT \cite{colagrande2025copift} and COPIFTv2 \cite{colagrande2026copiftv2}.
Pseudo-dual issue showed that a simple single-issue frontend can incrementally construct a superscalar schedule across loop iterations, but only for trivial, fully separable loop schedules.
COPIFT and COPIFTv2 relax this restriction, but do so by shifting communication and synchronization back to software, resulting in a more complex and less transparent programming model.
\glspl{slp} retain the key insight behind this line of work, namely the loop-local reuse of an execution schedule across iterations, while replacing explicit software-managed inter-thread communication with hardware dependency tracking.

\emph{Conventional OoO superscalar execution.} %
In \Cref{sec:architecture} we compared \glspl{slp} with conventional \gls{ooo} superscalar processors.
\glspl{slp} borrow several core mechanisms from conventional designs, including dependence tracking \cite{tomasulo1967, yeager1996, kessler1999}, register renaming \cite{tomasulo1967, yeager1996, kessler1999}, and physical-register reclamation \cite{yeager1996, sohi1995, moudgill1993, akkary2003, battle2012}, but use them to construct a schedule only during the first iteration of a loop.
This is the key distinction from general-purpose \gls{ooo} execution: an \gls{ooo} core continuously solves a dynamic scheduling problem for arbitrary code, whereas \glspl{slp} specialize for regular loops and amortize schedule construction across iterations by leveraging dataflow graph invariance.
This specialization further allows the issue width used during dataflow construction to be decoupled from the issue width realized during execution, substantially simplifying the scheduling logic.
This comes at the cost of reduced generality, as \glspl{slp} trade the flexibility of general-purpose \gls{ooo} execution for lower scheduling complexity and improved area and energy efficiency.

\emph{Loop specialization in OoO processors.} %
%
Loop-oriented specializations have also been explored in superscalar designs.
Loop caches \cite{kin1997, bellas1999, lea1999, anderson2000} exploit the fact that hot loops repeatedly execute the same instructions, reducing instruction-cache access, fetch, branch prediction, and (in some designs) decode work by caching and replaying instructions or decoded micro-ops for small loops.

Revolver \cite{hayenga2014revolver} moves this idea closer to the backend: it keeps two unrolled copies of a loop resident in an \gls{ooo} core's issue queues, from which they are replayed to further eliminate repeated rename, allocate and dispatch work.
Revolver and \glspl{slp} therefore share the idea of caching loop execution state beyond the frontend.
In this sense, Revolver can be viewed as realizing the \gls{slp} principle as an energy-saving mode within a conventional \gls{ooo} processor, while still relying on the costly machinery of general-purpose \gls{ooo} execution.
Standalone \glspl{slp}, by contrast, realize this principle in a loop-specialized architecture, enabling superscalar execution in lightweight in-order \glspl{pe} without a conventional \gls{ooo} backend.
Thus, Revolver preserves the flexibility of general-purpose \gls{ooo} processors while reaping the energy-efficiency benefits of the \gls{slp} paradigm, whereas standalone \glspl{slp} trade off this flexibility to further realize the \gls{slp} paradigm's area-efficiency benefits.
This difference is also reflected in their cached state: Revolver stores two unrolled loop bodies and assigns two physical destination registers to each result-producing loop instruction, whereas Schnizo stores a single loop body in its \glspl{rs} and reuses a single physical register across iterations.
This distinction is important at the lightweight \gls{pe} design point, critical for accelerator integration, targeted by Schnizo, where \gls{rs} and \gls{prf} storage are primary area contributors.

Finally, LoopFrog \cite{erdos2025loopfrog} leverages loop regularity by spawning future loop iterations as speculative ``threadlets'' inside a wide \gls{ooo} core.
While their approach targets parallelism across loop iterations, our approach exposes \gls{ilp} within a loop iteration, making the two approaches largely orthogonal.
Moreover, while LoopFrog handles inter-iteration dependencies speculatively, squashing threadlets upon ordering violations, \glspl{slp} support loop-carried dependencies directly through explicit dependency recording.

\emph{Statically-scheduled superscalar architectures.} %
%
Other low-complexity alternatives to \gls{ooo} superscalar execution include architectures that shift scheduling from hardware to software, such as VLIW \cite{fisher1983, colwell1987}, EDGE \cite{sankaralingam2006, burger2004}, and TTA-style processors \cite{corporaal1994}.
Like \glspl{slp}, these architectures aim to exploit \gls{ilp} without incurring the full hardware cost of dynamic scheduling.
The key difference is where the schedule resides.
In VLIW, EDGE and TTA-style architectures, much of the execution schedule and communication structure is determined statically by the compiler and encoded in the program representation, which must therefore reflect the target machine organization.
In \glspl{slp}, by contrast, software still presents an ordinary scalar loop body, and the hardware derives a loop-local schedule dynamically during the first loop iteration.
This removes the need for compiler support to derive and encode implementation-specific schedules in the binary, thereby improving portability across implementations.

\emph{Vector processors.} %
Vector processors are a natural comparison point for \glspl{slp} as both architectures target regular loop kernels, yet do so through different abstractions.
Vector programs are typically derived from loop programs through loop vectorization, whereas \glspl{slp} treat scalar loops as first-class citizens.
Thus, although both architectures target the same class of programs, the vector and loop abstractions lead to concrete differences in scheduling, live state granularity, and the way \gls{ilp} is exposed.
Specifically, vector processors expose \gls{ilp} by chaining dependent vector operations within the pipelined vector datapath \cite{russell1978, asanovic1998}, but cannot overlap the scalar operations in the surrounding stripmining loop, such as reductions, scalar recurrences, or address-generation logic.
These operations have been observed to become a bottleneck with short vectors \cite{villa1998, lazo2022}, which are commonly employed in resource-constrained and massively-parallel configurations \cite{lazo2022, perotti2025spatz}.
Short vectors are attractive precisely because vector processors track live state at vector granularity: increasing the vector length increases vector \gls{rf} capacity requirements and, in turn, hardware cost.
\glspl{slp}, by contrast, track live state at vector-element granularity.
Therefore, their state requirements do not scale with vector length or, equivalently, with the number of loop iterations.
To this end, coupling an \gls{slp} that can capture \gls{ilp} uniformly across all stripmining loop operations to a vector processor that can efficiently extract \gls{dlp} could represent a promising approach.
We leave a detailed exploration of this design point, as well as a quantitative comparison against vector processors, to future work.

\section{Conclusion}

In this work, we introduced \acrfullpl{slp}, a lightweight superscalar architecture specialized for regular, compute-intensive loop kernels.
Unlike conventional \gls{ooo} superscalar processors, \glspl{slp} solve the dynamic scheduling problem only in the first iteration of a loop and reuse the resulting dependence information in subsequent iterations.
By decoupling the issue width used for schedule construction from the issue width realized during execution, \glspl{slp} can achieve multiple-issue execution without incurring the full complexity of dynamic \gls{ooo} scheduling.
While standalone \glspl{slp} restrict superscalar execution to regular loops, the \gls{slp} paradigm could be realized as a specialized execution mode in a general-purpose \gls{ooo} core, preserving support for arbitrary code while possibly improving energy efficiency on hot loops.

We presented Schnizo, a working prototype of the \gls{slp} architecture developed as an enhancement to a tiny RV32I core.
We release Schnizo as open-source hardware to support future architectural research on \glspl{slp}.
Through its design, we highlighted both the general challenges and the practical solutions involved in building effective \glspl{slp}.
At the same time, Schnizo represents only one point in the \gls{slp} design space.
Alternative implementations based on a centralized physical \gls{rf}, more aggressive pipelining, or shared constant memories across \glspl{fu} remain open for exploration together with their associated tradeoffs.

We evaluated the scalability of Schnizo, showing that the degree of parallelism can be efficiently scaled, whereas extending support to larger loop bodies incurs storage costs that grow linearly.
Although our proof-of-concept implementation is built entirely from \glspl{ff}, all structures expose low port counts and therefore lend themselves to more area-efficient SRAM-based implementations.
Such implementations could mitigate the cost of scaling to larger loop bodies, both in the sequential storage elements and the mux-heavy combinational port logic.

On a diverse set of loop microkernels, Schnizo delivers substantial improvements in \gls{ipc} and \gls{fpu} utilization over the scalar baseline, reaching up to a geomean \ResultGeomeanFractionOfIdealIpcExclDiv\% of the ideal \gls{ipc} and \gls{fpu} utilization attainable on an ideal architecture with the same degree of parallelism.
On simple linear algebra kernels, a specialized Schnizo configuration achieves a geomean speedup of \ResultLaGeomeanSpeedup$\times$ over scalar execution while improving core area efficiency by \ResultLaGeomeanAreaEfficiencyImpr\%.

Overall, these results show that \glspl{slp} offer a compelling path for programmable many-core accelerators targeting regular, compute-intensive workloads, delivering higher \gls{fu} utilization while preserving the efficiency advantages of lightweight cores that are key to scaling performance.

\ifdefined\blindreview
\else
\section*{Acknowledgements}
This work has received funding from the Swiss State Secretariat for Education, Research, and Innovation (SERI) under the SwissChips initiative.
\fi


\appendix[Independence of Performance on Loop-Body Size]
\label{sec:appendix-slp-proof}

Given a loop with $L$ loop-body instructions and $N$ iterations, attaining \gls{ipc} of $\text{IPC}_\text{DCP}$ and $\text{IPC}_\text{DEP}$, respectively, in \gls{dcp} and \gls{dep} mode, the following equations hold, by definition of \gls{ipc}:
\begin{align*}
  \text{instructions}_\text{DCP} &= 2 \cdot L \\
  \text{instructions}_\text{DEP} &= (N-2) \cdot L \\
  \text{cycles}_\text{DCP} &= 2 \cdot L / \text{IPC}_\text{DCP} \\
  \text{cycles}_\text{DEP} &= (N-2) \cdot L / \text{IPC}_\text{DEP}
\end{align*}

Again, by definition of \gls{ipc}, the overall \gls{ipc} of the loop is given by:
\begin{align*}
  \text{IPC}_\text{overall} &= \frac{\text{instructions}}{\text{cycles}} = \\
  &= \frac{\text{instructions}_\text{DCP} + \text{instructions}_\text{DEP}}{\text{cycles}_\text{DCP} + \text{cycles}_\text{DEP}} = \\
  &= \frac{2 \cdot L + (N-2) \cdot L}{2 \cdot L / \text{IPC}_\text{DCP} + (N-2) \cdot L / \text{IPC}_\text{DEP}} = \\
  &= \frac{N}{2 / \text{IPC}_\text{DCP} + (N-2) / \text{IPC}_\text{DEP}} = \\
  &= \frac{\text{IPC}_\text{DCP} \cdot \text{IPC}_\text{DEP} \cdot N}{2 \cdot \text{IPC}_\text{DEP} + (N-2) \cdot \text{IPC}_\text{DCP}} = \\
  &= \frac{\text{IPC}_\text{DCP} \cdot \text{IPC}_\text{DEP} \cdot N}{\text{IPC}_\text{DCP} \cdot N + 2 \cdot \text{IPC}_\text{DEP} - 2 \cdot \text{IPC}_\text{DCP}} = \\
  &= \frac{a \cdot N}{b \cdot N + c}
\end{align*}

This formula reflects the empirically-validated curves in \Cref{fig:axpy_perf}, where $a$, $b$, and $c$ are given by:
\begin{align*}
  a &= \text{IPC}_\text{DCP} \cdot \text{IPC}_\text{DEP} \\
  b &= \text{IPC}_\text{DCP} \\
  c &= 2 \cdot (\text{IPC}_\text{DEP} - \text{IPC}_\text{DCP})
\end{align*}
That is, $a$, $b$, and $c$ are constants that do not depend on $L$, and the overall \gls{ipc} is only a function of the number of iterations $N$.

\ifdefined\blindreview
\bibliographystyle{ACM-Reference-Format}
\else
\bibliographystyle{IEEEtran}
\fi
\bibliography{paper}

@article{openai_compute,
  title={AI and Compute},
  author={Amodei, Dario and Hernandez, Danny},
  journal={OpenAI Blog},
  year={2018},
  url={https://openai.com/research/ai-and-compute}
}

@techreport{openai_enterprise_ai_2025,
  author       = {{OpenAI}},
  title        = {The State of Enterprise AI},
  year         = {2025},
  institution  = {OpenAI},
  url          = {https://cdn.openai.com/pdf/7ef17d82-96bf-4dd1-9df2-228f7f377a29/the-state-of-enterprise-ai_2025-report.pdf}
}

@report{hai2025_ai_index,
  title        = {AI Index Report 2025},
  author       = {{Stanford Institute for Human-Centered Artificial Intelligence}},
  institution  = {Stanford HAI},
  year         = {2025},
  url          = {https://hai.stanford.edu/ai-index/2025-ai-index-report}
}

@article{kaplan2020scaling,
  author       = {Jared Kaplan and Sam McCandlish and Tom Henighan and Tom B. Brown and Benjamin Chess and Rewon Child and Scott Gray and Alec Radford and Jeffrey Wu and Dario Amodei},
  title        = {Scaling Laws for Neural Language Models},
  journal      = {arXiv preprint arXiv:2001.08361},
  year         = {2020},
  doi          = {10.48550/arXiv.2001.08361},
  url          = {https://arxiv.org/abs/2001.08361}
}

@inproceedings{hoffmann2022compute,
  author       = {Jordan Hoffmann and Sebastian Borgeaud and Arthur Mensch and Elena Buchatskaya and Trevor Cai and Eliza Rutherford and Diego de Las Casas and Lisa Anne Hendricks and Johannes Welbl and Aidan Clark and Thomas Hennigan and Eric Noland and Katherine Millican and George van den Driessche and Bogdan Damoc and Aurelia Guy and Simon Osindero and Kar{\'e}n Simonyan and Erich Elsen and Oriol Vinyals and Jack W. Rae and Laurent Sifre},
  title        = {An Empirical Analysis of Compute-Optimal Large Language Model Training},
  booktitle    = {Advances in Neural Information Processing Systems},
  volume       = {35},
  pages        = {30016--30030},
  year         = {2022},
  url          = {https://proceedings.neurips.cc/paper_files/paper/2022/file/c1e2faff6f588870935f114ebe04a3e5-Paper-Conference.pdf}
}

@techreport{sevilla2024training,
  author       = {Jaime Sevilla and Edu Rold{\'a}n},
  title        = {Training Compute of Frontier {AI} Models Grows by 4--5x per Year},
  institution  = {Epoch AI},
  year         = {2024},
  month        = may,
  url          = {https://epoch.ai/blog/training-compute-of-frontier-ai-models-grows-by-4-5x-per-year}
}

@inproceedings{jung2024hammerblade,
  author       = {Dai Cheol Jung and Max Ruttenberg and Paul Gao and Scott Davidson and Daniel Petrisko and Kangli Li and Aditya K. Kamath and Lin Cheng and Shaolin Xie and Peitian Pan and Zhongyuan Zhao and Zichao Yue and Bandhav Veluri and Sripathi Muralitharan and Adrian Sampson and Andrew Lumsdaine and Zhiru Zhang and Christopher Batten and Mark Oskin and Dustin Richmond and Michael Bedford Taylor},
  title        = {Scalable, Programmable and Dense: The HammerBlade Open-Source {RISC-V} Manycore},
  booktitle    = {Proceedings of the Annual International Symposium on Computer Architecture (ISCA)},
  pages        = {770--784},
  year         = {2024},
  publisher    = {IEEE},
  doi          = {10.1109/ISCA59077.2024.00061},
  url          = {https://doi.org/10.1109/ISCA59077.2024.00061}
}

@inproceedings{abts2020thinkfast,
  author       = {Dennis Abts and Jonathan Ross and Jonathan Sparling and Mark Wong-VanHaren and Max Baker and Tom Hawkins and Andrew Bell and John Thompson and Temesghen Kahsai and Garrin Kimmell and Jennifer Hwang and Rebekah Leslie-Hurd and Michael Bye and E. R. Creswick and Matthew Boyd and Mahitha Venigalla and Evan Laforge and Jon Purdy and Purushotham Kamath and Dinesh Maheshwari and Michael Beidler and Geert Rosseel and Omar Ahmad and Gleb Gagarin and Richard Czekalski and Ashay Rane and Sahil Parmar and Jeff Werner and Jim Sproch and Adri{\'a}n Mac{\'i}as and Brian Kurtz},
  title        = {Think Fast: A Tensor Streaming Processor ({TSP}) for Accelerating Deep Learning Workloads},
  booktitle    = {Proceedings of the Annual International Symposium on Computer Architecture (ISCA)},
  pages        = {145--158},
  year         = {2020},
  publisher    = {IEEE},
  doi          = {10.1109/ISCA45697.2020.00023},
  url          = {https://doi.org/10.1109/ISCA45697.2020.00023}
}

@article{zaruba2021snitch,
  author       = {Florian Zaruba and Fabian Schuiki and Torsten Hoefler and Luca Benini},
  title        = {Snitch: A Tiny Pseudo Dual-Issue Processor for Area and Energy Efficient Execution of Floating-Point Intensive Workloads},
  journal      = {IEEE Transactions on Computers},
  volume       = {70},
  number       = {11},
  pages        = {1845--1860},
  year         = {2021},
  doi          = {10.1109/TC.2020.3027900},
  url          = {https://doi.org/10.1109/TC.2020.3027900}
}

@INPROCEEDINGS{colagrande2025copift,
  author={Colagrande, Luca and Benini, Luca},
  booktitle={2025 62nd ACM/IEEE Design Automation Conference (DAC)}, 
  title={Dual-Issue Execution of Mixed Integer and Floating-Point Workloads on Energy-Efficient In-Order RISC-V Cores}, 
  year={2025},
  volume={},
  number={},
  pages={1-7},
  doi={10.1109/DAC63849.2025.11132520}
}

@INPROCEEDINGS{colagrande2026copiftv2,
  author={Colagrande, Luca and Benini, Luca},
  booktitle={2026 Design, Automation \& Test in Europe Conference (DATE)}, 
  title={Late Breaking Results: Boosting Efficient Dual-Issue Execution on Lightweight RISC-V Cores}, 
  year={2026},
  volume={},
  number={},
  pages={1-3},
  doi={10.23919/DATE69613.2026.11539067}}

@ARTICLE{tomasulo1967,
  author={Tomasulo, R. M.},
  journal={IBM Journal of Research and Development}, 
  title={An Efficient Algorithm for Exploiting Multiple Arithmetic Units}, 
  year={1967},
  volume={11},
  number={1},
  pages={25-33},
  doi={10.1147/rd.111.0025}
}

@book{allen2001,
  author    = {Randy Allen and Ken Kennedy},
  title     = {Optimizing Compilers for Modern Architectures: A Dependence-based Approach},
  publisher = {Morgan Kaufmann},
  year      = {2001},
  isbn      = {9781558602861}
}

@book{aho2006,
  author    = {Alfred V. Aho and Monica S. Lam and Ravi Sethi and Jeffrey D. Ullman},
  title     = {Compilers: Principles, Techniques, and Tools},
  edition   = {2},
  publisher = {Pearson},
  year      = {2006},
  isbn      = {9780321486813}
}

@inproceedings{hind2001,
author = {Hind, Michael},
title = {Pointer analysis: haven't we solved this problem yet?},
year = {2001},
isbn = {1581134134},
publisher = {Association for Computing Machinery},
address = {New York, NY, USA},
url = {https://doi.org/10.1145/379605.379665},
doi = {10.1145/379605.379665},
booktitle = {Proceedings of the 2001 ACM SIGPLAN-SIGSOFT Workshop on Program Analysis for Software Tools and Engineering},
pages = {54–61},
numpages = {8},
location = {Snowbird, Utah, USA},
series = {PASTE '01}
}

@book{banerjee1997,
  author    = {Utpal Banerjee},
  title     = {Dependence Analysis},
  publisher = {Kluwer Academic Publishers},
  year      = {1997},
  isbn      = {9780792397335}
}

@inproceedings{amdahl1967,
author = {Amdahl, Gene M.},
title = {Validity of the single processor approach to achieving large scale computing capabilities},
year = {1967},
isbn = {9781450378956},
publisher = {Association for Computing Machinery},
address = {New York, NY, USA},
url = {https://doi.org/10.1145/1465482.1465560},
doi = {10.1145/1465482.1465560},
booktitle = {Proceedings of the April 18-20, 1967, Spring Joint Computer Conference},
pages = {483–485},
numpages = {3},
location = {Atlantic City, New Jersey},
series = {AFIPS '67 (Spring)}
}

@inproceedings{fu2025,
  author    = {Zexin Fu and Riccardo Tedeschi and Gianmarco Ottavi and
               Nils Wistoff and C{\'e}sar Fuguet and Davide Rossi and Luca Benini},
  title     = {Ramping Up Open-Source RISC-V Cores: Assessing the Energy Efficiency
               of Superscalar, Out-of-Order Execution},
  booktitle = {Proceedings of the 22nd ACM International Conference on Computing Frontiers (CF '25)},
  pages     = {12--20},
  year      = {2025},
  publisher = {ACM},
}

@INPROCEEDINGS{c910,
  author={Chen, Chen and Xiang, Xiaoyan and Liu, Chang and Shang, Yunhai and Guo, Ren and Liu, Dongqi and Lu, Yimin and Hao, Ziyi and Luo, Jiahui and Chen, Zhijian and Li, Chunqiang and Pu, Yu and Meng, Jianyi and Yan, Xiaolang and Xie, Yuan and Qi, Xiaoning},
  booktitle={2020 ACM/IEEE 47th Annual International Symposium on Computer Architecture (ISCA)}, 
  title={Xuantie-910: A Commercial Multi-Core 12-Stage Pipeline Out-of-Order 64-bit High Performance RISC-V Processor with Vector Extension : Industrial Product}, 
  year={2020},
  volume={},
  number={},
  pages={52-64},
  doi={10.1109/ISCA45697.2020.00016}}

@inproceedings{tanasic2017,
author = {Tanasic, Ivan and Gelado, Isaac and Jorda, Marc and Ayguade, Eduard and Navarro, Nacho},
title = {Efficient exception handling support for GPUs},
year = {2017},
isbn = {9781450349529},
publisher = {Association for Computing Machinery},
address = {New York, NY, USA},
url = {https://doi.org/10.1145/3123939.3123950},
doi = {10.1145/3123939.3123950},
booktitle = {Proceedings of the 50th Annual IEEE/ACM International Symposium on Microarchitecture},
pages = {109–122},
numpages = {14},
location = {Cambridge, Massachusetts},
series = {MICRO-50 '17}
}

@inproceedings{menon2012igpu,
author = {Menon, Jaikrishnan and De Kruijf, Marc and Sankaralingam, Karthikeyan},
title = {iGPU: exception support and speculative execution on GPUs},
year = {2012},
isbn = {9781450316422},
publisher = {IEEE Computer Society},
address = {USA},
booktitle = {Proceedings of the 39th Annual International Symposium on Computer Architecture},
pages = {72–83},
numpages = {12},
location = {Portland, Oregon},
series = {ISCA '12}
}

@inproceedings{li2023gpufpx,
author = {Li, Xinyi and Laguna, Ignacio and Fang, Bo and Swirydowicz, Katarzyna and Li, Ang and Gopalakrishnan, Ganesh},
title = {Design and Evaluation of GPU-FPX: A Low-Overhead tool for Floating-Point Exception Detection in NVIDIA GPUs},
year = {2023},
isbn = {9798400701559},
publisher = {Association for Computing Machinery},
address = {New York, NY, USA},
url = {https://doi.org/10.1145/3588195.3592991},
doi = {10.1145/3588195.3592991},
booktitle = {Proceedings of the 32nd International Symposium on High-Performance Parallel and Distributed Computing},
pages = {59–71},
numpages = {13},
location = {Orlando, FL, USA},
series = {HPDC '23}
}

@manual{amd_rocgdb_memory,
  title        = {ROCgdb Documentation: AMD GPU Memory Violation Reporting},
  author       = {{Advanced Micro Devices, Inc.}},
  organization = {AMD},
  year         = {2026},
  note         = {Accessed 2026-04-04},
  url          = {https://rocm.docs.amd.com/projects/ROCgdb/en/amd-staging/ROCgdb/gdb/doc/gdb/AMD-GPU.html#amd-gpu-memory-violation-reporting}
}

@manual{amd_rocgdb_alu,
  title        = {ROCgdb Documentation: AMD GPU ALU Exception Reporting},
  author       = {{Advanced Micro Devices, Inc.}},
  organization = {AMD},
  year         = {2026},
  note         = {Accessed 2026-04-04},
  url          = {https://rocm.docs.amd.com/projects/ROCgdb/en/amd-staging/ROCgdb/gdb/doc/gdb/AMD-GPU.html#amd-gpu-alu-exception-reporting}
}

@manual{nvidia_cuda_gdb_errors,
  title        = {CUDA-GDB Documentation: GPU Error Reporting},
  author       = {{NVIDIA Corporation}},
  organization = {NVIDIA},
  year         = {2026},
  note         = {Accessed 2026-04-04},
  url          = {https://docs.nvidia.com/cuda/cuda-gdb/index.html#gpu-error-reporting}
}

@ARTICLE{iliakis2022,
  author={Iliakis, Konstantinos and Xydis, Sotirios and Soudris, Dimitrios},
  journal={IEEE Transactions on Parallel and Distributed Systems}, 
  title={Repurposing GPU Microarchitectures with Light-Weight Out-Of-Order Execution}, 
  year={2022},
  volume={33},
  number={2},
  pages={388-402},
  doi={10.1109/TPDS.2021.3093231}}

@inproceedings{firoozshahian2023mtia,
author = {Firoozshahian, Amin and Coburn, Joel and Levenstein, Roman and Nattoji, Rakesh and Kamath, Ashwin and Wu, Olivia and Grewal, Gurdeepak and Aepala, Harish and Jakka, Bhasker and Dreyer, Bob and Hutchin, Adam and Diril, Utku and Nair, Krishnakumar and Aredestani, Ehsan K. and Schatz, Martin and Hao, Yuchen and Komuravelli, Rakesh and Ho, Kunming and Abu Asal, Sameer and Shajrawi, Joe and Quinn, Kevin and Sreedhara, Nagesh and Kansal, Pankaj and Wei, Willie and Jayaraman, Dheepak and Cheng, Linda and Chopda, Pritam and Wang, Eric and Bikumandla, Ajay and Karthik Sengottuvel, Arun and Thottempudi, Krishna and Narasimha, Ashwin and Dodds, Brian and Gao, Cao and Zhang, Jiyuan and Al-Sanabani, Mohammed and Zehtabioskuie, Ana and Fix, Jordan and Yu, Hangchen and Li, Richard and Gondkar, Kaustubh and Montgomery, Jack and Tsai, Mike and Dwarakapuram, Saritha and Desai, Sanjay and Avidan, Nili and Ramani, Poorvaja and Narayanan, Karthik and Mathews, Ajit and Gopal, Sethu and Naumov, Maxim and Rao, Vijay and Noru, Krishna and Reddy, Harikrishna and Venkatapuram, Prahlad and Bjorlin, Alexis},
title = {MTIA: First Generation Silicon Targeting Meta's Recommendation Systems},
year = {2023},
isbn = {9798400700958},
publisher = {Association for Computing Machinery},
address = {New York, NY, USA},
url = {https://doi.org/10.1145/3579371.3589348},
doi = {10.1145/3579371.3589348},
booktitle = {Proceedings of the 50th Annual International Symposium on Computer Architecture},
articleno = {80},
numpages = {13},
location = {Orlando, FL, USA},
series = {ISCA '23}
}

@ARTICLE{choquette2023blackwell,
  author={Choquette, Jack},
  journal={IEEE Micro},
  title={{NVIDIA Hopper H100 GPU: Scaling Performance}},
  year={2023},
  volume={43},
  number={3},
  pages={9-17},
}

@inproceedings{coburn2025mtia2,
author = {Coburn, Joel and Tang, Chunqiang and Asal, Sameer Abu and Agrawal, Neeraj and Chinta, Raviteja and Dixit, Harish and Dodds, Brian and Dwarakapuram, Saritha and Firoozshahian, Amin and Gao, Cao and Gondkar, Kaustubh and Graf, Tyler and Hu, Junhan and Huang, Jian and Hughes, Sterling and Hutchin, Adam and Jakka, Bhasker and Chen, Guoqiang Jerry and Kalyanaraman, Indu and Kamath, Ashwin and Kansal, Pankaj and Kazi, Erum and Levenstein, Roman and Maddury, Mahesh and Mastro, Alex and Medaiyese, Siji and Modi, Pritesh and Montgomery, Jack and Nadathur, Satish and Nagpal, Amit and Narasimha, Ashwin and Naumov, Maxim and Ozer, Eleanor and Park, Jongsoo and Ramani, Poorvaja and Reddy, Harikrishna and Reiss, David and Roy, Deboleena and Sekar, Sathish and Sharma, Arushi and Shetty, Pavan and Sukumaran-Rajam, Aravind and Tal, Eran and Tsai, Mike and Varshini, Shreya and Wareing, Richard and Wu, Olivia and Xie, Xiaolong and Yang, Jinghan and Yu, Hangchen and Zargar, Tanmay and Zeng, Zitong and Zhang, Feixiong and Matthews, Ajit and Jiao, Xun and Zhang, Jiyuan and Menage, Emmanuel and Stokke, Truls Edvard and Sourouri, Mohammed},
title = {Meta's Second Generation AI Chip: Model-Chip Co-Design and Productionization Experiences},
year = {2025},
isbn = {9798400712616},
publisher = {Association for Computing Machinery},
address = {New York, NY, USA},
url = {https://doi.org/10.1145/3695053.3731409},
doi = {10.1145/3695053.3731409},
booktitle = {Proceedings of the 52nd Annual International Symposium on Computer Architecture},
pages = {1689–1702},
numpages = {14},
location = {
},
series = {ISCA '25}
}

@inproceedings{prabhakar2024sn40l,
author = {Prabhakar, Raghu and Sivaramakrishnan, Ram and Gandhi, Darshan and Du, Yun and Wang, Mingran and Song, Xiangyu and Zhang, Kejie and Gao, Tianren and Wang, Angela and Li, Xiaoyan and Sheng, Yongning and Brot, Joshua and Sokolov, Denis and Vivek, Apurv and Leung, Calvin and Sabnis, Arjun and Bai, Jiayu and Zhao, Tuowen and Gottscho, Mark and Jackson, David and Luttrell, Mark and Shah, Manish K. and Chen, Zhengyu and Liang, Kaizhao and Jain, Swayambhoo and Thakker, Urmish and Huang, Dawei and Jairath, Sumti and Brown, Kevin J. and Olukotun, Kunle},
title = {SambaNova SN40L: Scaling the AI Memory Wall with Dataflow and Composition of Experts},
year = {2024},
publisher = {IEEE Press},
url = {https://doi.org/10.1109/MICRO61859.2024.00100},
doi = {10.1109/MICRO61859.2024.00100},
booktitle = {Proceedings of the 2024 57th IEEE/ACM International Symposium on Microarchitecture},
pages = {1353–1366},
numpages = {14},
location = {Austin, TX, USA},
series = {MICRO '24}
}

@INPROCEEDINGS{vasiljevic2024blackhole,
  author={Vasiljevic, Jasmina and Capalija, Davor},
  booktitle={2024 IEEE Hot Chips 36 Symposium (HCS)}, 
  title={{Blackhole} \& {TT-Metalium}: The Standalone {AI} Computer and its Programming Model}, 
  year={2024},
  volume={},
  number={},
  pages={1-30},
}

@inproceedings{lie2024wse3,
  author    = {Sean Lie},
  title     = {Wafer-Scale {AI}: {GPU} Impossible Performance},
  booktitle = {Proceedings of the 36th {IEEE} Hot Chips Symposium (HCS)},
  year      = {2024},
  pages     = {1--71},
  publisher = {IEEE}
}

@article{rico2024xdna,
  author  = {Alejandro Rico and Satyaprakash Pareek and Javier Cabezas and David Clarke and Baris Ozgul and Francisco Barat and Yao Fu and Stephan M{\"u}nz and Dylan Stuart and Patrick Schlangen and Pedro Duarte and Sneha Date and Indrani Paul and Jian Weng and Sonal Santan and Vinod Kathail and Ashish Sirasao and Juanjo Noguera},
  title   = {{AMD XDNA}\texttrademark\ {NPU} in Ryzen\texttrademark\ {AI} Processors},
  journal = {IEEE Micro},
  volume  = {44},
  number  = {6},
  year    = {2024},
  pages   = {62--71},
  doi     = {10.1109/MM.2024.3423692}
}

@ARTICLE{scheffler2025occamy,
  author={Scheffler, Paul and Benz, Thomas and Potocnik, Viviane and Fischer, Tim and Colagrande, Luca and Wistoff, Nils and Zhang, Yichao and Bertaccini, Luca and Ottavi, Gianmarco and Eggimann, Manuel and Cavalcante, Matheus and Paulin, Gianna and Gürkaynak, Frank K. and Rossi, Davide and Benini, Luca},
  journal={IEEE Journal of Solid-State Circuits}, 
  title={Occamy: A 432-Core Dual-Chiplet Dual-HBM2E 768-DP-GFLOP/s RISC-V System for 8-to-64-bit Dense and Sparse Computing in 12-nm FinFET}, 
  year={2025},
  volume={60},
  number={4},
  pages={1324-1338},
  doi={10.1109/JSSC.2025.3529249}}

@INPROCEEDINGS{liao2021ascend,
  author={Liao, Heng and Tu, Jiajin and Xia, Jing and Liu, Hu and Zhou, Xiping and Yuan, Honghui and Hu, Yuxing},
  booktitle={2021 IEEE International Symposium on High-Performance Computer Architecture (HPCA)}, 
  title={Ascend: a Scalable and Unified Architecture for Ubiquitous Deep Neural Network Computing : Industry Track Paper}, 
  year={2021},
  volume={},
  number={},
  pages={789-801},
  doi={10.1109/HPCA51647.2021.00071}}

@ARTICLE{baas2007asap,
  author={Baas, Bevan and Yu, Zhiyi and Meeuwsen, Michael and Sattari, Omar and Apperson, Ryan and Work, Eric and Webb, Jeremy and Lai, Michael and Mohsenin, Tinoosh and Truong, Dean and Cheung, Jason},
  journal={IEEE Micro}, 
  title={AsAP: A Fine-Grained Many-Core Platform for DSP Applications}, 
  year={2007},
  volume={27},
  number={2},
  pages={34-45},
  doi={10.1109/MM.2007.29}}

@INPROCEEDINGS{bohnenstiehl2016,
  author={Bohnenstiehl, Brent and Stillmaker, Aaron and Pimentel, Jon and Andreas, Timothy and Bin Liu and Tran, Anh and Adeagbo, Emmanuel and Baas, Bevan},
  booktitle={2016 IEEE Symposium on VLSI Circuits (VLSI-Circuits)}, 
  title={A 5.8 pJ/Op 115 billion ops/sec, to 1.78 trillion ops/sec 32nm 1000-processor array}, 
  year={2016},
  volume={},
  number={},
  pages={1-2},
  doi={10.1109/VLSIC.2016.7573511}}

@inproceedings{miyajima2025,
author = {Miyajima, Takaaki and Fukuoka, Leon},
title = {Benchmarking the Cerebras Wafer Scale Engine-2 Architecture},
year = {2025},
isbn = {9798400718717},
publisher = {Association for Computing Machinery},
address = {New York, NY, USA},
url = {https://doi.org/10.1145/3731599.3767438},
doi = {10.1145/3731599.3767438},
booktitle = {Proceedings of the SC '25 Workshops of the International Conference for High Performance Computing, Networking, Storage and Analysis},
pages = {818–822},
numpages = {5},
location = {
},
series = {SC Workshops '25}
}

@article{palacharla1997complexity,
  title={Complexity-Effective Superscalar Processors},
  author={Palacharla, S. and Jouppi, Norman P. and Smith, James E.},
  journal={Proceedings of the 24th Annual International Symposium on Computer Architecture (ISCA)},
  year={1997}
}

@ARTICLE{kucuk2004,
  author={Kucuk, G. and Ponomarev, D.V. and Ergin, O. and Ghose, K.},
  journal={IEEE Transactions on Computers}, 
  title={Complexity-effective reorder buffer designs for superscalar processors}, 
  year={2004},
  volume={53},
  number={6},
  pages={653-665},
  doi={10.1109/TC.2004.5}}

@inproceedings{folegnani2001,
author = {Folegnani, Daniele and Gonz\'{a}lez, Antonio},
title = {Energy-effective issue logic},
year = {2001},
isbn = {0769511627},
publisher = {Association for Computing Machinery},
address = {New York, NY, USA},
url = {https://doi.org/10.1145/379240.379266},
doi = {10.1145/379240.379266},
booktitle = {Proceedings of the 28th Annual International Symposium on Computer Architecture},
pages = {230–239},
numpages = {10},
location = {G\"{o}teborg, Sweden},
series = {ISCA '01}
}

@INPROCEEDINGS{huang2002,
  author={Huang, M. and Renau, J. and Torrellas, J.},
  booktitle={Proceedings of the International Symposium on Low Power Electronics and Design}, 
  title={Energy-efficient hybrid wakeup logic}, 
  year={2002},
  volume={},
  number={},
  pages={196-201},
  doi={10.1109/LPE.2002.146736}}

@INPROCEEDINGS{karami2025nongemm,
  author={Karami, Rachid and Kao, Sheng-Chun and Kwon, Hyoukjun},
  booktitle={2025 IEEE International Symposium on Performance Analysis of Systems and Software (ISPASS)}, 
  title={Understanding the Performance Horizon of the Latest ML Workloads with NonGEMM Workloads}, 
  year={2025},
  volume={},
  number={},
  pages={1-14},
  doi={10.1109/ISPASS64960.2025.00011}}

@inproceedings{feng2023tensorir,
  author       = {Siyuan Feng and Bohan Hou and Hongyi Jin and Wuwei Lin and Junru Shao and Ruihang Lai and Zihao Ye and Lianmin Zheng and Cody Hao Yu and Yong Yu and Tianqi Chen},
  title        = {TensorIR: An Abstraction for Automatic Tensorized Program Optimization},
  booktitle    = {Proceedings of the ACM International Conference on Architectural Support for Programming Languages and Operating Systems (ASPLOS)},
  pages        = {804--817},
  year         = {2023},
  publisher    = {Association for Computing Machinery},
  doi          = {10.1145/3575693.3576933},
  url          = {https://doi.org/10.1145/3575693.3576933}
}

@misc{vasilache2018,
      title={Tensor Comprehensions: Framework-Agnostic High-Performance Machine Learning Abstractions}, 
      author={Nicolas Vasilache and Oleksandr Zinenko and Theodoros Theodoridis and Priya Goyal and Zachary DeVito and William S. Moses and Sven Verdoolaege and Andrew Adams and Albert Cohen},
      year={2018},
      eprint={1802.04730},
      archivePrefix={arXiv},
      primaryClass={cs.PL},
      url={https://arxiv.org/abs/1802.04730}, 
}

@ARTICLE{gautschi2017pulp,
  author={Gautschi, Michael and Schiavone, Pasquale Davide and Traber, Andreas and Loi, Igor and Pullini, Antonio and Rossi, Davide and Flamand, Eric and Gürkaynak, Frank K. and Benini, Luca},
  journal={IEEE Transactions on Very Large Scale Integration (VLSI) Systems}, 
  title={Near-Threshold RISC-V Core With DSP Extensions for Scalable IoT Endpoint Devices}, 
  year={2017},
  volume={25},
  number={10},
  pages={2700-2713},
  doi={10.1109/TVLSI.2017.2654506}
}

@ARTICLE{cui2023,
  author={Cui, Enfang and Li, Tianzheng and Wei, Qian},
  journal={IEEE Access}, 
  title={RISC-V Instruction Set Architecture Extensions: A Survey}, 
  year={2023},
  volume={11},
  number={},
  pages={24696-24711},
  doi={10.1109/ACCESS.2023.3246491}}

@article{galuzzi2011,
author = {Galuzzi, Carlo and Bertels, Koen},
title = {The Instruction-Set Extension Problem: A Survey},
year = {2011},
issue_date = {May 2011},
publisher = {Association for Computing Machinery},
address = {New York, NY, USA},
volume = {4},
number = {2},
issn = {1936-7406},
url = {https://doi.org/10.1145/1968502.1968509},
doi = {10.1145/1968502.1968509},
journal = {ACM Trans. Reconfigurable Technol. Syst.},
month = may,
articleno = {18},
numpages = {28}
}

@ARTICLE{talla2003,
  author={Talla, D. and John, L.K. and Burger, D.},
  journal={IEEE Transactions on Computers}, 
  title={Bottlenecks in multimedia processing with SIMD style extensions and architectural enhancements}, 
  year={2003},
  volume={52},
  number={8},
  pages={1015-1031},
}

@ARTICLE{kavvadias2008,
  author={Kavvadias, Nikolaos and Nikolaidis, Spiridon},
  journal={IEEE Transactions on Computers}, 
  title={Elimination of Overhead Operations in Complex Loop Structures for Embedded Microprocessors}, 
  year={2008},
  volume={57},
  number={2},
  pages={200-214},
}

@INPROCEEDINGS{kavvadias2010,
  author={Kavvadias, Nikolaos and Masselos, Konstantinos},
  booktitle={IEEE Computer Society Annual Symposium on VLSI (ISVLSI)}, 
  title={Efficient Hardware Looping Units for FPGAs}, 
  year={2010},
  volume={},
  number={},
  pages={35-40},
}

@ARTICLE{bajwa1997,
  author={Bajwa, R.S. and Hiraki, M. and Kojima, H. and Gorny, D.J. and Nitta, K. and Shridhar, A. and Seki, K. and Sasaki, K.},
  journal={IEEE Transactions on Very Large Scale Integration (VLSI) Systems}, 
  title={Instruction buffering to reduce power in processors for signal processing}, 
  year={1997},
  volume={5},
  number={4},
  pages={417-424},
  doi={10.1109/92.645068}
}

@inproceedings{uh1999,
author = {Uh, Gang-Ryung and Wang, Yuhong and Whalley, David and Jinturkar, Sanjay and Burns, Chris and Cao, Vincent},
title = {Effective exploitation of a zero overhead loop buffer},
year = {1999},
booktitle = {ACM SIGPLAN 1999 Workshop on Languages, Compilers, and Tools for Embedded Systems (LCTES)},
pages = {10–19},
}

@INPROCEEDINGS{tsao2003,
  author={Ya-Lan Tsao and Wei-Hao Chen and Wen-Sheng Cheng and Maw-Ching Lin and Shyh-Jye Jou},
  booktitle={IEEE International Systems-on-Chip Conference (SOC)}, 
  title={Hardware nested looping of parameterized and embedded DSP core}, 
  year={2003},
  volume={},
  number={},
  pages={49-52},
}

@INPROCEEDINGS{vadivel2017,
  author={Vadivel, Kanishkan and Wijtvliet, Mark and Jordans, Roel and Corporaal, Henk},
  booktitle={Euromicro Conference on Digital System Design (DSD)}, 
  title={Loop Overhead Reduction Techniques for Coarse Grained Reconfigurable Architectures}, 
  year={2017},
  volume={},
  number={},
  pages={14-21},
}

@INPROCEEDINGS{wei2010,
  author={Zhenqi Wei and Peilin Liu and Ji Kong and Rendong Ying},
  booktitle={International Symposium on Intelligent Signal Processing and Communication Systems (ISPACS)}, 
  title={Low-power microarchitecture of zero-overhead nested loops in embedded processors}, 
  year={2010},
  volume={},
  number={},
  pages={1-4},
}

@INPROCEEDINGS{colagrande2025gemm,
  author={Colagrande, Luca and Leone, Lorenzo and Coco, Maximilian and Deaconeasa, Andrei and Benini, Luca},
  booktitle={2025 IEEE/ACM International Symposium on Low Power Electronics and Design (ISLPED)}, 
  title={Towards Zero-Stall Matrix Multiplication on Energy-Efficient RISC-V Clusters for Machine Learning Acceleration}, 
  year={2025},
  volume={},
  number={},
  pages={1-7},
  doi={10.1109/ISLPED65674.2025.11261759}}

@inproceedings{fan2025,
author = {Fan, Enming and Guan, Xiaofeng and Hu, Fan and Shi, Heng and Zhou, Hao and Yao, Jianguo},
title = {Postiz: Extending Post-increment Addressing for Loop Optimization and Code Size Reduction},
year = {2025},
isbn = {9798400712753},
publisher = {Association for Computing Machinery},
address = {New York, NY, USA},
url = {https://doi.org/10.1145/3696443.3708935},
doi = {10.1145/3696443.3708935},
booktitle = {Proceedings of the 23rd ACM/IEEE International Symposium on Code Generation and Optimization},
pages = {600–613},
numpages = {14},
location = {Las Vegas, NV, USA},
series = {CGO '25}
}

@ARTICLE{lee1987,
  author={Lee, E. and Messerschmitt, D.},
  journal={IEEE Transactions on Acoustics, Speech, and Signal Processing}, 
  title={Pipeline interleaved programmable DSP's: Architecture}, 
  year={1987},
  volume={35},
  number={9},
  pages={1320-1333},
  doi={10.1109/TASSP.1987.1165274}}

@article{scheffler2023sssr,
author = {Scheffler, Paul and Zaruba, Florian and Schuiki, Fabian and Hoefler, Torsten and Benini, Luca},
title = {Sparse Stream Semantic Registers: A Lightweight ISA Extension Accelerating General Sparse Linear Algebra},
year = {2023},
issue_date = {Dec. 2023},
publisher = {IEEE Press},
volume = {34},
number = {12},
issn = {1045-9219},
url = {https://doi.org/10.1109/TPDS.2023.3322029},
doi = {10.1109/TPDS.2023.3322029},
journal = {IEEE Trans. Parallel Distrib. Syst.},
month = dec,
pages = {3147–3161},
numpages = {15}
}

@article{schuiki2021ssr,
  author       = {Fabian Schuiki and Florian Zaruba and Torsten Hoefler and Luca Benini},
  title        = {Stream Semantic Registers: A Lightweight {RISC-V} {ISA} Extension Achieving Full Compute Utilization in Single-Issue Cores},
  journal      = {IEEE Transactions on Computers},
  volume       = {70},
  number       = {2},
  pages        = {212--227},
  year         = {2021},
  doi          = {10.1109/TC.2020.2987314},
  url          = {https://doi.org/10.1109/TC.2020.2987314}
}

@article{yeager1996,
author = {Yeager, Kenneth C.},
title = {The MIPS R10000 Superscalar Microprocessor},
year = {1996},
issue_date = {April 1996},
publisher = {IEEE Computer Society Press},
address = {Washington, DC, USA},
volume = {16},
number = {2},
issn = {0272-1732},
url = {https://doi.org/10.1109/40.491460},
doi = {10.1109/40.491460},
journal = {IEEE Micro},
month = apr,
pages = {28–40},
numpages = {13}
}

@ARTICLE{kessler1999,
  author={Kessler, R.E.},
  journal={IEEE Micro}, 
  title={The Alpha 21264 microprocessor}, 
  year={1999},
  volume={19},
  number={2},
  pages={24-36},
  doi={10.1109/40.755465}}

@INPROCEEDINGS{balasubramonian2001,
  author={Balasubramonian, R. and Dwarkadas, S. and Albonesi, D.H.},
  booktitle={Proceedings. 34th ACM/IEEE International Symposium on Microarchitecture. MICRO-34}, 
  title={Reducing the complexity of the register file in dynamic superscalar processors}, 
  year={2001},
  volume={},
  number={},
  pages={237-248},
  doi={10.1109/MICRO.2001.991122}}

@INPROCEEDINGS{tseng2003,
  author={Tseng, J.H. and Asanovic, K.},
  booktitle={30th Annual International Symposium on Computer Architecture, 2003. Proceedings.}, 
  title={Banked multiported register files for high-frequency superscalar microprocessors}, 
  year={2003},
  volume={},
  number={},
  pages={62-71},
  doi={10.1109/ISCA.2003.1206989}}

@ARTICLE{sohi1995,
  author={Smith, J.E. and Sohi, G.S.},
  journal={Proceedings of the IEEE}, 
  title={The microarchitecture of superscalar processors}, 
  year={1995},
  volume={83},
  number={12},
  pages={1609-1624},
  doi={10.1109/5.476078}}

@inproceedings{moudgill1993,
author = {Moudgill, Mayan and Pingali, Keshav and Vassiliadis, Stamatis},
title = {Register renaming and dynamic speculation: an alternative approach},
year = {1993},
isbn = {0818652802},
publisher = {IEEE Computer Society Press},
address = {Washington, DC, USA},
booktitle = {Proceedings of the 26th Annual International Symposium on Microarchitecture},
pages = {202–213},
numpages = {12},
location = {Austin, Texas, USA},
series = {MICRO 26}
}

@INPROCEEDINGS{akkary2003,
  author={Akkary, H. and Rajwar, R. and Srinivasan, S.T.},
  booktitle={Proceedings. 36th Annual IEEE/ACM International Symposium on Microarchitecture, 2003. MICRO-36.}, 
  title={Checkpoint processing and recovery: towards scalable large instruction window processors}, 
  year={2003},
  volume={},
  number={},
  pages={423-434},
  doi={10.1109/MICRO.2003.1253246}}

@INPROCEEDINGS{battle2012,
  author={Battle, Steven and Hilton, Andrew D. and Hempstead, Mark and Roth, Amir},
  booktitle={IEEE International Symposium on High-Performance Computer Architecture}, 
  title={Flexible register management using reference counting}, 
  year={2012},
  volume={},
  number={},
  pages={1-12},
  doi={10.1109/HPCA.2012.6169033}}

@INPROCEEDINGS{farkas1997,
  author={Farkas, K.I. and Chow, P. and Jouppi, N.P. and Vranesic, Z.},
  booktitle={Proceedings of 30th Annual International Symposium on Microarchitecture}, 
  title={The multicluster architecture: reducing cycle time through partitioning}, 
  year={1997},
  volume={},
  number={},
  pages={149-159},
  doi={10.1109/MICRO.1997.645806}}

@article{sohi1995multiscalar,
author = {Sohi, Gurindar S. and Breach, Scott E. and Vijaykumar, T. N.},
title = {Multiscalar processors},
year = {1995},
issue_date = {May 1995},
publisher = {Association for Computing Machinery},
address = {New York, NY, USA},
volume = {23},
number = {2},
issn = {0163-5964},
url = {https://doi.org/10.1145/225830.224451},
doi = {10.1145/225830.224451},
journal = {SIGARCH Comput. Archit. News},
month = may,
pages = {414–425},
numpages = {12}
}

@article{smith2001,
author = {Smith, James E.},
title = {Instruction-Level Distributed Processing},
year = {2001},
issue_date = {April 2001},
publisher = {IEEE Computer Society Press},
address = {Washington, DC, USA},
volume = {34},
number = {4},
issn = {0018-9162},
url = {https://doi.org/10.1109/2.917541},
doi = {10.1109/2.917541},
journal = {Computer},
month = apr,
pages = {59–65},
numpages = {7}
}

@ARTICLE{taylor2005,
  author={Taylor, M.D. and Lee, W. and Amarasinghe, S.P. and Agarwal, A.},
  journal={IEEE Transactions on Parallel and Distributed Systems}, 
  title={Scalar operand networks}, 
  year={2005},
  volume={16},
  number={2},
  pages={145-162},
  doi={10.1109/TPDS.2005.24}}

@inproceedings{allen1983,
author = {Allen, J. R. and Kennedy, Ken and Porterfield, Carrie and Warren, Joe},
title = {Conversion of control dependence to data dependence},
year = {1983},
isbn = {0897910907},
publisher = {Association for Computing Machinery},
address = {New York, NY, USA},
url = {https://doi.org/10.1145/567067.567085},
doi = {10.1145/567067.567085},
booktitle = {Proceedings of the 10th ACM SIGACT-SIGPLAN Symposium on Principles of Programming Languages},
pages = {177–189},
numpages = {13},
location = {Austin, Texas},
series = {POPL '83}
}

@inproceedings{mahlke1992,
author = {Mahlke, Scott A. and Lin, David C. and Chen, William Y. and Hank, Richard E. and Bringmann, Roger A.},
title = {Effective compiler support for predicated execution using the hyperblock},
year = {1992},
isbn = {0818631759},
publisher = {IEEE Computer Society Press},
address = {Washington, DC, USA},
booktitle = {Proceedings of the 25th Annual International Symposium on Microarchitecture},
pages = {45–54},
numpages = {10},
location = {Portland, Oregon, USA},
series = {MICRO 25}
}

@INPROCEEDINGS{kennedy1990,
  author={Kennedy, K. and McKinley, K.S.},
  booktitle={Supercomputing '90:Proceedings of the 1990 ACM/IEEE Conference on Supercomputing}, 
  title={Loop distribution with arbitrary control flow}, 
  year={1990},
  volume={},
  number={},
  pages={407-416},
  doi={10.1109/SUPERC.1990.130048}}

@book{banerjee1993,
author = {Banerjee, Utpal K.},
title = {Loop Transformations for Restructuring Compilers: The Foundations},
year = {1993},
isbn = {079239318X},
publisher = {Kluwer Academic Publishers},
address = {USA}
}

@INPROCEEDINGS{kin1997,
  author={Kin, J. and Munish Gupta and Mangione-Smith, W.H.},
  booktitle={Proceedings of 30th Annual International Symposium on Microarchitecture}, 
  title={The filter cache: an energy efficient memory structure}, 
  year={1997},
  volume={},
  number={},
  pages={184-193},
  doi={10.1109/MICRO.1997.645809}}

@INPROCEEDINGS{bellas1999,
  author={Bellas, N. and Hajj, I. and Polychronopoulos, C. and Stamoulis, G.},
  booktitle={Proceedings 1999 IEEE International Conference on Computer Design: VLSI in Computers and Processors (Cat. No.99CB37040)}, 
  title={Energy and performance improvements in microprocessor design using a loop cache}, 
  year={1999},
  volume={},
  number={},
  pages={378-383},
  doi={10.1109/ICCD.1999.808570}}

@INPROCEEDINGS{lea1999,
  author={Lea Hwang Lee and Moyer, B. and Arends, J.},
  booktitle={Proceedings. 1999 International Symposium on Low Power Electronics and Design (Cat. No.99TH8477)}, 
  title={Instruction fetch energy reduction using loop caches for embedded applications with small tight loops}, 
  year={1999},
  volume={},
  number={},
  pages={267-269},
  doi={10.1109/LPE.1999.799454}}

@INPROCEEDINGS{anderson2000,
  author={Anderson, T. and Agarwala, S.},
  booktitle={Proceedings 2000 International Conference on Computer Design}, 
  title={Effective hardware-based two-way loop cache for high performance low power processors}, 
  year={2000},
  volume={},
  number={},
  pages={403-407},
  doi={10.1109/ICCD.2000.878315}}

@inproceedings{erdos2025loopfrog,
author = {Erdos, Marton and Bora, Utpal and Bhosale, Akshay and Lytton, Bob and Zaidi, Ali M. and Chadwick, Alexandra W. and Guo, Yuxin and Gabrielli, Giacomo and Jones, Timothy M.},
title = {LoopFrog: In-Core Hint-Based Loop Parallelization},
year = {2025},
isbn = {9798400715730},
publisher = {Association for Computing Machinery},
address = {New York, NY, USA},
url = {https://doi.org/10.1145/3725843.3756051},
doi = {10.1145/3725843.3756051},
booktitle = {Proceedings of the 58th IEEE/ACM International Symposium on Microarchitecture},
pages = {490–503},
numpages = {14},
location = {
},
series = {MICRO '25}
}

@article{fisher1983,
author = {Fisher, Joseph A.},
title = {Very Long Instruction Word architectures and the ELI-512},
year = {1983},
issue_date = {June 1983},
publisher = {Association for Computing Machinery},
address = {New York, NY, USA},
volume = {11},
number = {3},
issn = {0163-5964},
url = {https://doi.org/10.1145/1067651.801649},
doi = {10.1145/1067651.801649},
journal = {SIGARCH Comput. Archit. News},
month = jun,
pages = {140–150},
numpages = {11}
}

@INPROCEEDINGS{corporaal1994,
  author={Corporaal, H.},
  booktitle={Proceedings of 4th Great Lakes Symposium on VLSI}, 
  title={Design of transport triggered architectures}, 
  year={1994},
  volume={},
  number={},
  pages={130-135},
  doi={10.1109/GLSV.1994.289981}}

@INPROCEEDINGS{sankaralingam2006,
  author={Sankaralingam, Karthikeyan and Nagarajan, Ramadass and McDonald, Robert and Desikan, Rajagopalan and Drolia, Saurabh and Govindan, M.S. and Gratz, Paul and Gulati, Divya and Hanson, Heather and Kim, Changkyu and Liu, Haiming and Ranganathan, Nitya and Sethumadhavan, Simha and Sharif, Sadia and Shivakumar, Premkishore and Keckler, Stephen W. and Burger, Doug},
  booktitle={2006 39th Annual IEEE/ACM International Symposium on Microarchitecture (MICRO'06)}, 
  title={Distributed Microarchitectural Protocols in the TRIPS Prototype Processor}, 
  year={2006},
  volume={},
  number={},
  pages={480-491},
  doi={10.1109/MICRO.2006.19}}

@ARTICLE{burger2004,
  author={Burger, D. and Keckler, S.W. and McKinley, K.S. and Dahlin, M. and John, L.K. and Lin, C. and Moore, C.R. and Burrill, J. and McDonald, R.G. and Yoder, W.},
  journal={Computer}, 
  title={Scaling to the end of silicon with EDGE architectures}, 
  year={2004},
  volume={37},
  number={7},
  pages={44-55},
  doi={10.1109/MC.2004.65}}

@inproceedings{colwell1987,
author = {Colwell, Robert P. and Nix, Robert P. and O'Donnell, John J. and Papworth, David B. and Rodman, Paul K.},
title = {A VLIW architecture for a trace scheduling compiler},
year = {1987},
isbn = {0818608056},
publisher = {Association for Computing Machinery},
address = {New York, NY, USA},
url = {https://doi.org/10.1145/36206.36201},
doi = {10.1145/36206.36201},
booktitle = {Proceedings of the Second International Conference on Architectural Support for Programming Languages and Operating Systems},
pages = {180–192},
numpages = {13},
location = {Palo Alto, California, USA},
series = {ASPLOS II}
}

@inproceedings{villa1998,
author = {Villa, Luis and Espasa, Roger and Valero, Mateo},
title = {A performance study of out-of-order vector architectures and short registers},
year = {1998},
isbn = {089791998X},
publisher = {Association for Computing Machinery},
address = {New York, NY, USA},
url = {https://doi.org/10.1145/277830.277844},
doi = {10.1145/277830.277844},
booktitle = {Proceedings of the 12th International Conference on Supercomputing},
pages = {37–44},
numpages = {8},
location = {Melbourne, Australia},
series = {ICS '98}
}

@INPROCEEDINGS{lazo2022,
  author={Lazo, Cristóbal Ramírez and Reggiani, Enrico and Morales, Carlos Rojas and Figueras Bagué, Roger and Villa Vargas, Luis A. and Ramírez Salinas, Marco A. and Cortés, Mateo Valero and Sabri Ünsal, Osman and Cristal, Adrián},
  booktitle={2022 IEEE International Symposium on High-Performance Computer Architecture (HPCA)}, 
  title={Adaptable Register File Organization for Vector Processors}, 
  year={2022},
  volume={},
  number={},
  pages={786-799},
  doi={10.1109/HPCA53966.2022.00063}}

@ARTICLE{perotti2025spatz,
  author={Perotti, Matteo and Riedel, Samuel and Cavalcante, Matheus and Benini, Luca},
  journal={IEEE Transactions on Computer-Aided Design of Integrated Circuits and Systems}, 
  title={Spatz: Clustering Compact RISC-V-Based Vector Units to Maximize Computing Efficiency}, 
  year={2025},
  volume={44},
  number={7},
  pages={2488-2502},
  doi={10.1109/TCAD.2025.3528349}}

@article{russell1978,
author = {Russell, Richard M.},
title = {The CRAY-1 computer system},
year = {1978},
issue_date = {Jan. 1978},
publisher = {Association for Computing Machinery},
address = {New York, NY, USA},
volume = {21},
number = {1},
issn = {0001-0782},
url = {https://doi.org/10.1145/359327.359336},
doi = {10.1145/359327.359336},
journal = {Commun. ACM},
month = jan,
pages = {63–72},
numpages = {10}
}

@phdthesis{asanovic1998,
author = {Asanovic, Krste and Wawrzynek, John},
title = {Vector microprocessors},
year = {1998},
isbn = {0591990873},
publisher = {University of California, Berkeley},
note = {AAI9901978}
}

@inproceedings{dekruijf2011,
author = {de Kruijf, Marc and Sankaralingam, Karthikeyan},
title = {Idempotent processor architecture},
year = {2011},
isbn = {9781450310536},
publisher = {Association for Computing Machinery},
address = {New York, NY, USA},
url = {https://doi.org/10.1145/2155620.2155637},
doi = {10.1145/2155620.2155637},
booktitle = {Proceedings of the 44th Annual IEEE/ACM International Symposium on Microarchitecture},
pages = {140–151},
numpages = {12},
location = {Porto Alegre, Brazil},
series = {MICRO-44}
}

@INPROCEEDINGS{hayenga2014revolver,
  author={Hayenga, Mitchell and Naresh, Vignyan Reddy Kothinti and Lipasti, Mikko H.},
  booktitle={2014 IEEE 20th International Symposium on High Performance Computer Architecture (HPCA)}, 
  title={Revolver: Processor architecture for power efficient loop execution}, 
  year={2014},
  volume={},
  number={},
  pages={591-602},
  doi={10.1109/HPCA.2014.6835968}}

@INPROCEEDINGS{purayil2025troop,
  author={Purayil, Navaneeth Kunhi and Shen, Diyou and Perotti, Matteo and Benini, Luca},
  booktitle={2025 IEEE 43rd International Conference on Computer Design (ICCD)}, 
  title={TROOP: At-the-Roofline Performance for Vector Processors on Low Operational Intensity Workloads}, 
  year={2025},
  volume={},
  number={},
  pages={594-601},
  doi={10.1109/ICCD65941.2025.00091}}

\end{document}